%% file: Manuscript_arXiv_3.tex
\pdfoutput=1  % Add this as line 1 to force arXiv to use PDFLaTeX

\documentclass[american,aps,pra,reprint,superscriptaddress,longbibliography]{revtex4-1}
\usepackage[unicode=true,pdfusetitle, bookmarks=true,bookmarksnumbered=false,bookmarksopen=false, breaklinks=false,pdfborder={0 0 0},backref=false,colorlinks=false] {hyperref}
\hypersetup{colorlinks,linkcolor=myurlcolor,citecolor=myurlcolor,urlcolor=myurlcolor}
\usepackage{braket,colortbl,amsthm,amsmath,amssymb,txfonts,graphicx}
\definecolor{myurlcolor}{rgb}{0,0,0.7}
\usepackage{cleveref}
\usepackage{soul}
\usepackage{subfigure}

\usepackage{color}

\usepackage{pgffor} % <--- ADD THIS LINE HERE for the page loop macro

\begin{document}

\title{Quantum state localization in dipole-dipole interacting disordered networks}% atoms in high-Q cavities}
% \author{x}
% \email{y}
% \affiliation{z}

\author{Pritam Chattopadhyay}
\thanks{contributed equally}
% \email{pritam.chattopadhyay@weizmann.ac.il}
\affiliation{Department of Chemical and Biological Physics \& AMOS,
Weizmann Institute of Science, Rehovot 7610001, Israel}

\author{Saikat Sur}
\thanks{contributed equally}
% \email{saikat.sur@weizmann.ac.il}
\affiliation{Department of Chemical and Biological Physics \& AMOS,
Weizmann Institute of Science, Rehovot 7610001, Israel}
\affiliation{Optics \& Quantum Information Group, The Institute of Mathematical Sciences, HBNI, CIT Campus, Taramani, Chennai 600113, India}

\author{Avijit Misra}
% \thanks{contributed equally}
% \email{avijitmisra@iitism.ac.in}
% \affiliation{Department of Chemical and Biological Physics \& AMOS,
% Weizmann Institute of Science, Rehovot 7610001, Israel}
 \affiliation{Department of Physics, Indian Institute of Technology (ISM), Dhanbad, Jharkhand 826004, India}

\author{David Petrosyan}
% \email{dap@iesl.forth.gr}
\affiliation{Institute of Electronic Structure and Laser, FORTH, GR-70013 Heraklion, Greece}
\affiliation{Alikhanyan National Laboratory (YerPhI), 0036 Yerevan, Armenia}

\author{Arti Garg}
% \email{arti.garg@saha.ac.in}
\affiliation{Theory Division, Saha Institute of Nuclear Physics, 1/AF Bidhannagar, Kolkata 700 064, India}
\affiliation{Homi Bhabha National Institute, Training School Complex, Anushaktinagar, Mumbai 400094, India}

\author{Gershon Kurizki}
% \email{gershon.kurizki@weizmann.ac.il}
\affiliation{Department of Chemical and Biological Physics \& AMOS,
Weizmann Institute of Science, Rehovot 7610001, Israel}
 
\begin{abstract}

We study the localization of excitations in positionally disordered spin or atom networks coupled via the realistic resonant dipole–dipole interaction (RDDI), which does not conform to a simple power law, as the spatial dependence and dissipative character distinguish it from conventional short- or long-range models. Despite its partially long-ranged and radiative nature, positional disorder in the RDDI coupling leads to strong spatial localization of excitations. The interplay between coherent and dissipative couplings gives rise to nontrivial interference effects that stabilize localized modes even in open geometries. Our results uncover a photon wavelength-induced transition from extended to localized excitation dynamics, establishing RDDI networks as a unique setting to explore the emergence of localization in realistic quantum optical systems. Our analysis of the localized modes induced by RDDI  has potential applications in coherent photovoltaics, excitonic circuits, quantum memory, and quantum sensors.

\end{abstract}

\maketitle
\textit{Introduction:-}
In a generic isolated many-body quantum system, an initial non-equilibrium state gradually relaxes to a thermal-equilibrium state, a phenomenon commonly known as the eigenstate thermalization hypothesis (ETH)~\cite{d2016quantum,deutsch2018eigenstate,von1929proof,reimann2010canonical,rigol2008thermalization,srednicki1994chaos,cramer2008exact}. {Counter-examples to ETH have been many-body localization (MBL) phenomena in disordered multipartite interacting systems~\cite{vosk2015theory,shepelyansky_1994_prl,abanin_2019_rmp,basko2006metal,gornyi2005interacting,pal2010many,nandkishore2015many,alet2018many,abanin2019colloquium,sierant2025many,PhysRevResearch.3.033233,PhysRevX.7.041021}.} MBL provides a powerful tool for avoiding unwarranted transport and spreading of errors by isolating excitations at a particular local region and thereby realizing a resilient quantum memory~\cite{jing2023ensemble,wang2025entanglement,zhu2025remote,liu2025millisecond,moiseev2025optical,rui2015operating,julsgaard2013quantum}. The growing need for scalable and fault-tolerant quantum technologies, has given impetus to MBL investigations in diverse  (spin~\cite{santos2004entanglement,oganesyan2007localization,luitz2015many,aramthottil2024phenomenology}, bosonic~\cite{sierant2017many,sierant2018many,orell2019probing,hopjan2020many}, fermionic~\cite{mondaini2015many,prelovvsek2016absence,zakrzewski2018spin,kozarzewski2018spin,PhysRevB.99.224203,PhysRevB.96.060203,PhysRevB.109.214209}, and Floquet~\cite{lazarides2015fate,ponte2015many,ponte2015periodically,abanin2016theory,zhang2016floquet,bairey2017driving,sahay2021emergent,garratt2021many,sierant2023stability} ) systems.

While most MBL studies focus on diagonal disorder, i.e., randomness of on-site energies, \textit{off-diagonal disorder} (ODD) offers a less explored route to MBL, by giving rise to critical or non-exponentially localized phases in 1-D spin chains~\cite{abanin_2019_rmp,basko2006metal,abanin2019colloquium}. However, in these studies, ODD is mainly introduced in the \textit{short-ranged hopping} amplitude approximation~\cite{abanin_2019_rmp}. Furthermore, the robustness of ODD-induced MBL phases against \textit{dissipative effects} is largely unexplored, although dissipation may hamper MBL applications in fault-tolerant quantum technologies~\cite{Levi2016,Luschen2017}. 

Here, we study MBL \textit{solely driven by ODD} in hitherto unexplored spin or atom networks coupled by the resonant dipole-dipole interaction (RDDI). Such networks have eluded investigation despite their widespread presence in various platforms with dipole positional disorder. Because of the RDDI \textit{oscillatory} distance behavior, MBL behavior is found here to qualitatively differ from that of previously studied simplified MBL models that have been based on either \textit{nearest-neighbor}~\cite{shahmoon2016highly,lehmberg_pra_1970} or \textit{long-range interaction}~\cite{shahmoon2016highly,shahmoon2011strongly,PhysRevA.72.043803} (where the long-range interaction is assumed uniform). RDDI offers a realistic testbed for localization with a tunable range of interactions in the presence of dissipation, along with coherent exchange of excitation. A central hitherto unexplored result of our analysis is that random RDDI due to positional disorder leads to \textit{strong localization of both single- and multiple excitations}, despite the partially long-ranged character of the interaction and the associated collective radiative dissipation.

\begin{figure}[t]
    \centering
  \includegraphics[width=0.35\textwidth]{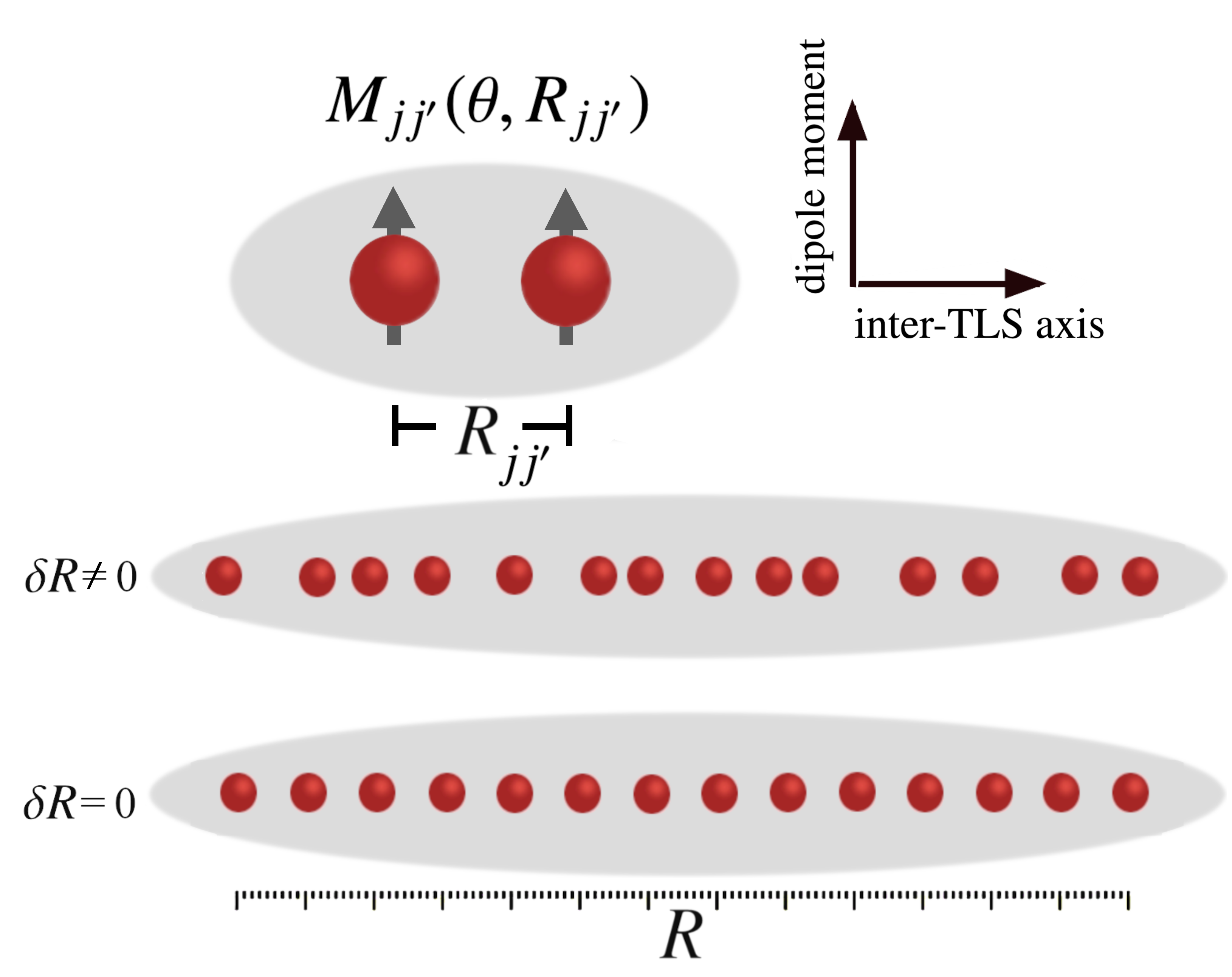}
% \includegraphics[width=0.55\textwidth]{prob_density_panel_2.pdf}
  % \subfigure[]{%
  % \includegraphics[width=0.45\textwidth]{spatial_distribution_t200_fixed.pdf}}
    \caption{{A schematic of ordered ($\delta R = 0$) and disordered  ($\delta R \ne 0$) chains of interacting atoms arranged in  one-dimension. Each atom is charge-neutral and associated with a dipole moment aligned perpendicularly with the inter-TLS axis. The average distance between two adjacent atoms is $R$ irrespective of the disorder present in the system.} }
    \label{Scheme}
\end{figure}

\textit{Model:-} We investigate a system of $N$ spatially distributed identical two-level systems (TLSs) (Fig.~\ref{Scheme}), realized by, e.g, Rydberg atoms~\cite{browaeys2016experimental}, superconducting qubits~\cite{wolfowicz2021quantum}, or electronic nitrogen vacancy (NV) spins in a diamond~\cite{doherty2013nitrogen}. The Hamiltonian for the  $N$-TLS RDDI-coupled system is 
\begin{subequations}
\begin{eqnarray}\label{ham}
 H = \sum^N_{j=1} \omega_0\vert e_j\rangle\mkern-2mu\langle e_j\vert  + \sum_{\substack{j,j'=1\\ j<j'}}^{N}M_{jj'} (\theta,R_{jj'}) (\vert e_j g_{j'} \rangle\mkern-2mu\langle  g_{j} e_{j'} \vert + \text{h.c.}).~~\end{eqnarray}
Here $\vert e_j \rangle$ and $\vert g_j \rangle$ are the excited and the ground states of the $j-$th TLS, respectively; $\omega_0$ is the dipolar transition frequency or the uniform two-level splitting caused by the transverse magnetic field.  The first term stands for the local energy of each excited TLS, while the second term describes coherent energy exchange between TLS pairs, $j$ and $j'$, \textit{not necessarily nearest neighbors}, via photon-mediated RDDI~\cite{thiru,lehmberg_pra_1970,ritter2012elementary,Chattopadhyay_2025,kofman_kurizki}, which is characterized by the coupling strength $M_{jj'}(\theta,R_{jj'}) $. %It is interesting to note that the long-range quadratic interaction between the TLSs can give rise to non-trivial strings of fermi creation and annihilation operators and therefore rendering features of many-body systems.
The RDDI between the  $j$~th and $j'$~th TLS has the full separation-dependent form  
% $M_{j_1j_2} (\theta,R_{j_1j_2}) = V_{j_1j_2}(\theta,R_{j_1j_2}) + i F_{j_1j_2} (\theta,R_{j_1j_2})$,
\begin{eqnarray}\label{EqM}
   && M_{jj'} (\theta,R_{jj'}) = V_{jj'}(\theta,R_{jj'}) + i F_{jj'} (\theta,R_{jj'}),
\end{eqnarray}
where $R_{jj'}$ denotes the separation between the atom pairs $j$ and $j'$, and $\theta$ is the angle between the dipole moment and the vector of the inter-TLS axis~\cite{thiru,defenu2023long,sheremet2023waveguide,ficek2005quantum,lehmberg_pra_1970}. %We assume the inter-TLS vector to be aligned perpendicularly for all TLS.} 
The real part
\begin{eqnarray}
     V_{jj'}(\theta,R_{jj'}) &=& \frac{\gamma}{4} \Bigg[(1-3\cos^2\theta) \left(\frac{\cos (k_0 R_{jj'})}{(k_0 R_{jj'})^3} + \frac{\sin(k_0 R_{jj'})}{(k_0 R_{jj'})^2}\right) \nonumber\\
  && - (1-  \cos^2\theta) \frac{\cos (k_0 R_{jj'})}{(k_0 R_{jj'} )} \Bigg];
\end{eqnarray}
represents coherent interaction, while the imaginary part 
\begin{eqnarray}\label{eqnnn4}
  F_{jj'}(\theta,R_{jj'}) &=& \frac{f \gamma}{4} \Bigg[(-1+3\cos^2\theta) \left(\frac{\sin (k_0 R_{jj'})}{(k_0 R_{jj'})^3} - \frac{\cos(k_0 R_{jj'})}{(k_0 R_{jj'})^2}\right) \nonumber\\
  && +(1-  \cos^2\theta) \frac{\sin (k_0 R_{jj'})}{(k_0 R_{jj'} )} \Bigg].\nonumber\\
\end{eqnarray}
\end{subequations}
accounts for collective radiative dissipation.

Here $\gamma$ is the single-TLS radiative decay rate at $\omega_0$, $k_0 =\omega_0/c =2\pi/\lambda_0$, the resonant photon wavelength $\lambda_0$ being the characteristic length scale of the RDDI. In a one-dimensional chain, $R_{jj'} = |X_{j} - X_{j'}|$, but generalization to other array geometries is straightforward. The parameter $f$ in Eq.~\eqref{eqnnn4} controls the strength of collective dissipation in the system: it is 1 in open space~\cite{lehmberg_pra_1970} but may be reduced to $0<f<1$ in photonic structures with modified density of states~\cite{shahmoon2016highly,shahmoon2011strongly,PhysRevA.72.043803, PhysRevA.42.2915}.
%and is generally set unity to describe realistic systems.}
Although the interaction in the atomic basis is pairwise, upon applying the Jordan–Wigner transformation~\cite{jordan_zfp_1928}, this model generates not only four-fermion interactions, but also higher-order fermion terms, leading to genuinely many-body behavior in the fermion basis (See SI). 

A distinct feature of the RDDI is its non-monotonic spatial dependence. If the distance between the adjacent TLS is much smaller than the resonant photon wavelength $\lambda_0 = 2\pi/k_0$, the real part of the RDDI scales as $V_{jj'} \propto (\lambda_0/|X_j-X_{j'}|)^3$, whereas its imaginary part scales as $F_{jj'} \propto (\lambda_0/|X_j-X_{j'}|)^2$. This scaling with distance implies that radiative dissipation becomes negligible compared to the coherent interaction as TLS become closely packed. Conversely, when the TLS are farther apart than the characteristic wavelength, both the real part and the imaginary parts of the RDDI scale as $V_{jj'} \sim F_{jj'}  \propto (\lambda_0/|X_j-X_{j'}|)$. Consequently, dissipation cannot be neglected, unless the TLS are confined within a high-Q cavity~\cite{paternostro2009solitonic,corzo2019waveguide,bardroff1995dynamical,PhysRevA.88.033830} or a photonic bandgap structure~\cite{shahmoon2011strongly,PhysRevA.72.043803,kofman1994spontaneous}.

\textit{Single-excitation localization via positional disorder:-} In the single-excitation subspace where one excitation is coherently shared among the TLSs, each excited TLS is coupled to all its unexcited counterparts via RDDI, but due to the nontrivial range dependence of this interaction, the scenario does not resemble standard all-to-all models where the interaction is equal among all partners~\cite{abanin_2019_rmp,abanin2019colloquium}.

For a long chain with perfectly ordered atoms with open boundary conditions, the bulk approximately respects translational invariance. Although edges explicitly break translational symmetry, the edge effect is less significant compared to that of the bulk, particularly in one dimension. {The disorder considered here is structural rather than energetic, as we assume that the atomic spatial coordinates fluctuate around an ordered reference array.} Such disorder breaks the symmetry of the system and allows localization, as shown below. Namely, the TLS positions $X_j$ in the translationally invariant network are randomly displaced by $\delta X_j$, which are drawn independently from a Gaussian distribution with zero mean value and standard deviation $\Delta_X$. We assume the weak disorder regime $\Delta_X << \vert X_j - X_{j+1}\vert$, so that the displacements only mildly change the TLS atomic positions, but the mean distance between two neighbors remains the same, i.e., $\langle \vert X_j - X_{j+1}\vert \rangle = R$. The randomized TLS separations $R_{jj'} =\vert X_{j}+ \delta X_{j} -X_{j'}- \delta X_{j'}\vert$, induce fluctuations in the RDDI terms $M_{jj'} (\theta, R_{jj'})$ in Eq.~\eqref{EqM}. Unlike toy models where the effect of disorder enters the Hamiltonian linearly~\cite{abanin_2019_rmp}, in RDDI the positional disorder modifies the pair separations $R_{jj'}$ that are arguments of oscillatory and long-range functions. This results in \textit{nonlinear corrections} (in powers of the displacements $\delta X_j$) in the disordered Hamiltonian, reflecting the nontrivial RDDI distance dependence. {This form of disorder is naturally realized in atomic, molecular, and
solid-state emitter arrays due to finite positioning accuracy, residual
motional spread, and structural inhomogeneity~\cite{PhysRevB.106.134212,PhysRevLett.117.020401,PhysRevLett.113.243002}. Importantly, unlike standard Anderson models with independent on-site disorder, here a single displacement $\delta X_j$
simultaneously modifies all couplings connected to site $j$. Thus, the
randomness is spatially correlated bond disorder generated by the
nonlinear and oscillatory RDDI kernel, not a phenomenological random
on-site potential.}
%of the dipole–dipole interaction.}

\begin{figure}
  \begin{center}
  \includegraphics[width=0.5\textwidth]{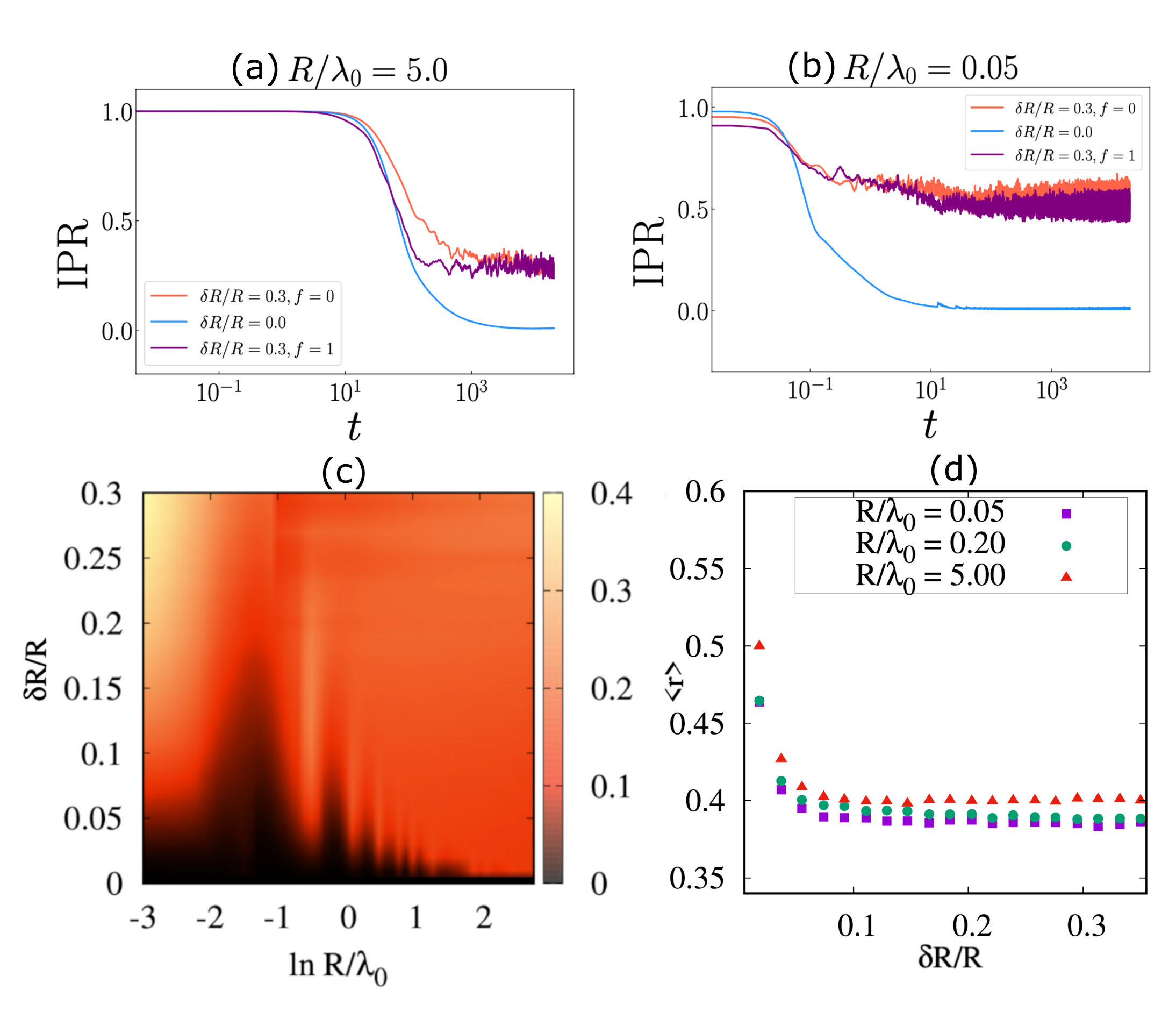}
  %        \subfigure[]{%
  % \includegraphics[width=0.46\textwidth]{IPR_t_rLamda0d5_1.pdf}}
  % \subfigure[]{%
  % \includegraphics[width=0.46\textwidth]{IPR_t_rLamda0d05_1.pdf}}
  % \subfigure[]{%
  %  \includegraphics[width=0.48\textwidth]{fig_static_IPR_f0.eps}}
  % \subfigure[]{%
  % \includegraphics[width=0.48\textwidth]{avg_r_f0.eps}}
  \end{center}
      \caption{ Dynamic IPR as a function of $t$ (in units of $\gamma^{-1}$) with and without disorder in $M$ for the (a) near-zone and (b) far-zone RDDI in the presence ($f=1$) and absence ($f=0$) of dissipation. {The initial state of the combined field-atom system in the \textit{single-excitation sector} is, $\vert \psi(0) \rangle = \sum_{j=2}^N \vert 1,g_j \rangle$.}  (c)  The static IPR for the off-diagonal disorder averaged over all eigenvectors with $\log(R/\lambda_0)$ for $N = 1000$ without dissipation for $f=0$. The color plot describes the variation of IPR for different randomness strengths, and (d) the average value of level spacing ratio $r$  as a function of disorder for different inter-atomic distances computed from the one excitation sector eigenvectors.  The plots are generated after averaging the data from two hundred independent disorder configurations. }
      \label{Prob_WdisM}
  \end{figure}

%\subsection{Inverse Participation Ratio}

\begin{figure}
  \begin{center}
\includegraphics[width=0.45\textwidth]{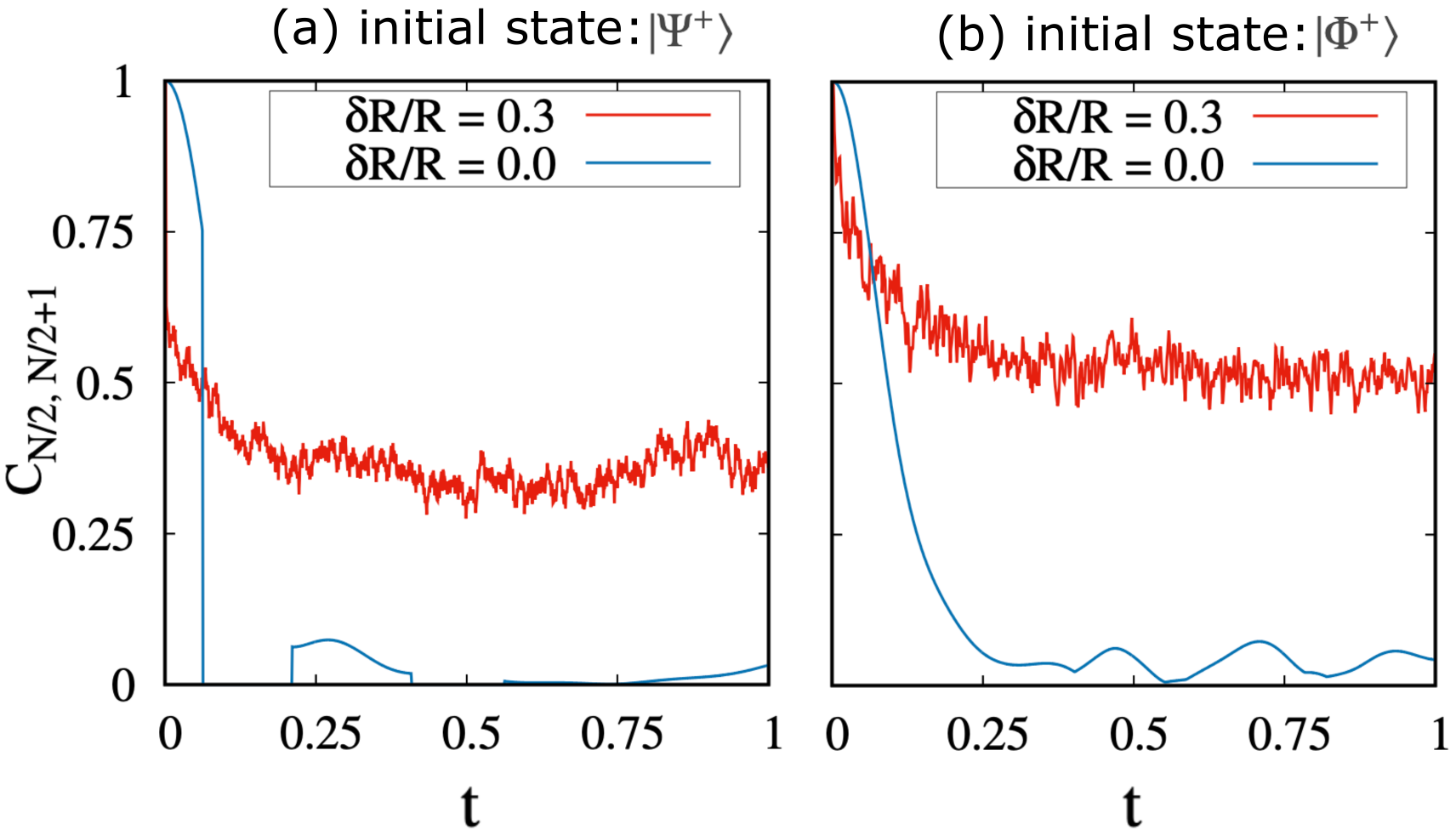}
  \end{center}
      \caption{The concurrence between $N/2~$th and $(N/2+1)~$th atom as a function of time (f=1). The system is initiated in a Bell state at the middle of the chain. (a)  A Bell state in a linear superposition of single excitations  $| \Psi^+\rangle =\frac{1}{\sqrt{2}}(\prod_{j \ne {N/2}}|e_{N/2}, g_j\rangle + \prod_{j \ne {N/2+1}}| e_{N/2+1}, g_j\rangle)$. (b) A Bell state in a linear superposition of zero and two excitations $| \Phi^+\rangle =\frac{1}{\sqrt{2}}(\prod_{j}|g_j\rangle + \prod_{j \ne {N/2},{N/2+1}}|e_{N/2}, e_{N/2+1}, g_j \rangle)$. The plots are generated by averaging the data from fifty independent disorder configurations.}
      \label{fidelity}
  \end{figure}

  \begin{figure}
  \begin{center}
\includegraphics[width=0.45\textwidth]{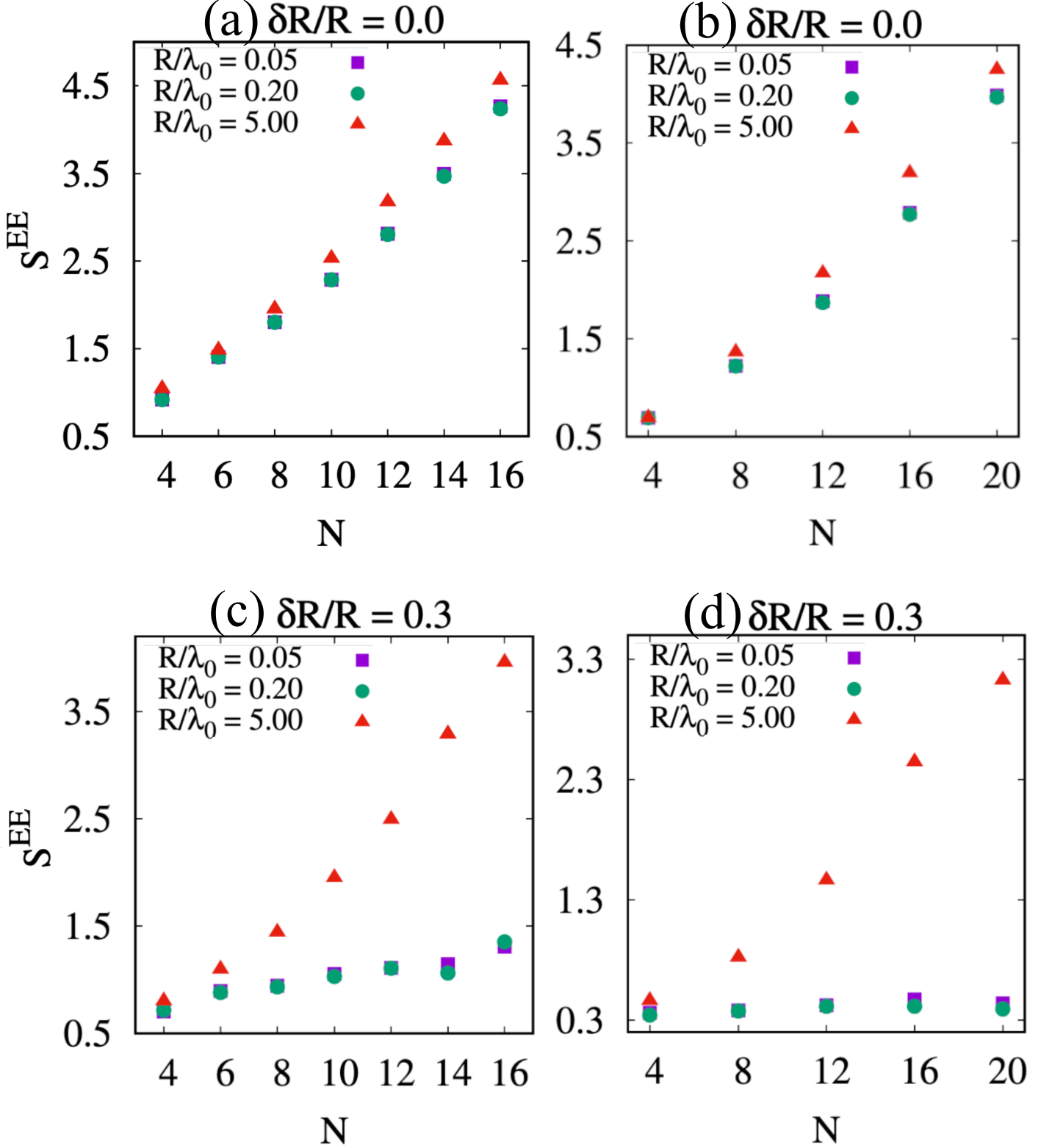}
  \end{center}
      \caption{The bipartite entanglement entropy averaged over all eigenstates as a function of system size: (a) without disorder $\delta R /R = 0.0$ for the excitation sector $N/2$, (b)  without disorder $\delta R /R = 0.0$ for the excitation sector $N/4$, (c)  with disorder $\delta R /R = 0.3$ for the excitation sector $N/2$, (d)  with disorder $\delta R /R = 0.3$ for the excitation sector $N/4$. The plots are generated after averaging the data from a hundred independent disorder configurations. }
      \label{fig:ee}
  \end{figure}

The most general time-dependent state of the $N$-TLS system in the \textit{single-excitation sector} is $\vert \psi(t) \rangle = \sum^N_{j=1}  \prod_{i\neq j} \alpha_j (t) \vert e_j,g_i\rangle$,
where $\vert e_j \rangle$ denotes a single-atom state in which the $j$-th TLS is excited, while the remaining ones are in their single-atom ground states $|g_i\rangle$. The complex, time-dependent  amplitudes
$\alpha_j (t)$  $ = \prod_{j \neq i}\langle e_j, g_i\vert\psi(t) \rangle$ %\st{are the wavefunctions that} 
evolve according to the Schr\"odinger equation,  $\ket{\dot{\psi}(t)}=-iH \ket{\psi(t)}$.

To quantitatively assess the localization of a many-body system, we employ two complementary measures: One is the well-established \textit{dynamic inverse participation ratio} (IPR)~\cite{edwards1972numerical},
\begin{eqnarray} \label{iprrr}
 {\color{black}IPR(\psi (t))= \sum^N_{j=1} \vert \alpha_j (t) \vert^4.}
% \st{\sum^N_{j=1} \prod_{i\neq j}|\langle e_j, g_i\vert  \psi (t) \rangle|^4 }  .    
\end{eqnarray}
The IPR measures the spatial concentration of the wavefunction across the basis states, with larger values indicating stronger localization.

The evolution of the dynamic IPR (Fig.~\ref{Prob_WdisM}) exhibits localization enhancement with increasing disorder. We plot the IPR for two representative values of the average TLS separation, $R$, and analyze the localization of excitation at the initially-excited site. Notably, the dynamic IPR asymptotically saturates at approximately $0.3$ in the far zone $R/\lambda_0 = 5.0$, while it doubles the saturation value to about 0.6 in the near zone $R/\lambda_0 = 0.05$, which signifies a substantially stronger localization. This behavior stems from the dominance of strong nearest-neighbor interactions in the near-zone limit, which inhibits its spatial delocalization across the TLS chain/array.

The \textit{static} IPR evaluates the same expression as \eqref{iprrr} for the energy eigenstates of the Hamiltonian, upon replacing $|\psi (t) \rangle$ by the stationary eigenstates, thereby providing insight into the localization of the system’s eigenmodes.

Figure~\ref{Prob_WdisM}(c) illustrates the variation of the static IPR as a function of $\ln(R/\lambda_0)$, under varying degrees of positional disorder (see Appendix B for the dissipative case). This static IPR also increases monotonically with disorder strength, signifying that positional randomness enhances the spatial confinement of the eigenstates, a hallmark of localization. This trend persists across the entire range of TLS separations, suggesting that disorder-induced localization is a robust feature of the system. Furthermore, the scaling behavior of the IPR with system size corroborates our identification of the transition from the localized to the extended regime (see Appendix C for detailed plots and discussion).

Closer inspection reveals three physically distinct regimes:\\
(i) In the near-zone regime $\ln(R/\lambda_0) < -1.5$, ($R<<\lambda_0$), the IPR increases steeply with disorder. Here, strong near-field RDDI dominate, enabling highly localized excitation modes that are confined to just a few TLS. The sharp IPR rise reflects the fact that even small positional deviations strongly affect the near-zone interference patterns governing the eigenstate structure.\\
(ii) In the far-zone regime, $\ln(R/\lambda_0) > -1.5$, where the TLS spacing exceeds the RDDI length scale, the dipolar coupling weakens, resulting in broader and less localized states. Although disorder still promotes localization, its influence is mitigated by the reduced interaction strength, yielding lower IPR values compared to the near-zone regime.\\

% }\\
(iii) Around $\ln(R/\lambda_0) \approx -1.5$, the system enters a transitional regime. Here, the IPR shows weaker sensitivity to disorder, indicating \textit{a crossover from the strongly localized near-zone} regime dominated by RDDI to the far-zone regime where disorder and spatial separation jointly suppress localization.

{\color{black}
A weak enhancement of the static IPR for $\ln(R/\lambda_0)>1$ is observed (Fig.~\ref{Prob_WdisM}(c)). This behavior originates from the retarded far-field form of the RDDI. In this regime, the oscillatory factors~\cite{PhysRevA.2.883} become highly sensitive to positional fluctuations so that even small fluctuations in $R_{jj'}$ generate random signs and phases in the long-range hopping amplitudes. This suppresses coherent transport through destructive interference and the IPR.
}

Across all regimes, the \textit{non-dissipative} system ($f=0$ in Eq.~\eqref{eqnnn4}) exhibits consistently higher IPR values than the dissipative one ($f=1$ in Eq.~\eqref{eqnnn4}) under identical conditions, because dissipation suppresses quantum coherence and, consequently, the interference necessary for strong localization. In the absence of dissipation, the system retains full coherence, allowing disorder to generate pronounced interference effects and tightly confined eigenstates.

Though a perturbative analysis of the disorder term cannot really prove localization of eigenstates, for an intuitive understanding, the analysis of the single-excitation manifold, induced by weak positional disorder under RDDI, is carried out (see SI). It shows that the eigenstates of the disordered system become spatially localized due to destructive interference of partial waves, consistent with our numerical analysis of the eigenstates.

%While the IPR provides both dynamical and eigenstate-based evidence of localization, it is instructive 
To corroborate the IPR-based findings, we examine the average level-spacing ratio $\langle r \rangle$ as a function of positional disorder (Fig.~\ref{Prob_WdisM}(d)). As disorder increases, $\langle r \rangle$ decreases and saturates near the Poissonian limit $2 \ln 2 - 1 \approx 0.386$, irrespective of the TLS separation. Among the three representative separations, the sharpest decay in $\langle r \rangle$ is observed for $R/\lambda_0 = 0.05$, precisely the regime where both the dynamic and static IPR indicate strong localization. This agreement between dynamical, eigenstate, and spectral diagnostics provides a consistent picture of the disorder-induced transition from extended to localized phases.

We also consider a \textit{spectral probe} based on the level-spacing ratio (LSR). In contrast to IPR, LSR is a measure of localization in equilibrium. %In the localized phase, the spacing between adjacent energy levels follows a Poisson distribution, whereas in the extended phase it obeys Gaussian Orthogonal Ensemble (GOE) statistics.
Denoting the ordered eigenvalues by ${E_n}$, we define the adjacent level spacings $\delta_n = E_{n+1} - E_n$ and introduce the dimensionless ratio
\begin{eqnarray}
  0 \le r_n = \frac{\text{min}\{\delta_n, \delta_{n+1}\}}{\text{max}\{\delta_n, \delta_{n+1}\}}  \le 1  
\end{eqnarray}
The statistical distribution of $r_n$ serves as a sensitive indicator of the underlying phase: Poisson statistics %\st{is a signal of } 
corresponds to localization, while the Gaussian orthogonal ensemble (Wigner-Dyson statistics) characterizes extended states.

Taken together, the IPR and LSR provide a consistent picture of localization in the $N$-TLS system. The IPR reveals how a \textit{single excitation} dynamically spreads or remains trapped in real space, while the LSR captures the statistical structure of the Hamiltonian spectrum. %These two diagnostic measures can establish a robust identification of localized and replace the extended waves as a function of disorder strengths.}

{\color{black}
A true single-particle mobility edge would require an energy $E_c$ separating localized and extended eigenstates in the thermodynamic limit. Operationally, this must be tested by energy-resolved finite-size scaling, for example, through $\mathrm{IPR}(E)=\left\langle \sum_{j=1}^{N}|\psi_E(j)|^4\right\rangle_{\rm dis}$. The IPR and level-spacing ratio shown in Fig.~\ref{Prob_WdisM} are averaged over the spectrum and therefore diagnose the global localization tendency but do not by themselves establish an energy-resolved mobility edge. Within the system sizes studied here, we do not observe a clear
separation into localized and extended energy windows. We therefore interpret the data as evidence for disorder-induced localization with energy-dependent localization lengths, rather than as evidence for a sharp single-particle mobility edge (see Appendix C, SI for details).

}

% \textit{Perturbative Analysis of Localization in the Single-Excitation Sector:}

\textit{Bell pair localization:-}
For quantum information applications, we study the dynamics of a Bell state formed by two TLS in the middle of the chain, $| \Psi^+\rangle =\frac{1}{\sqrt{2}}(\prod_{j \ne {N/2}}|e_{N/2}, g_j\rangle + \prod_{j \ne {N/2+1}}| e_{N/2+1}, g_j\rangle)$ and $| \Phi^+\rangle =\frac{1}{\sqrt{2}}(\prod_{j}|g_j\rangle + \prod_{j \ne {N/2},{N/2+1}}|e_{N/2}, e_{N/2+1}, g_j \rangle)$. The entanglement of the pair of atoms $N/2$ and $N/2+1$ in a randomly perturbed time-evolved quantum state is computed from their concurrence  $C_{N/2,N/2+1}(t)$~\cite{wootters}.
% \begin{eqnarray}
% \mathcal{F}(t) = \vert \langle \psi (0) \vert \psi(t) \rangle \vert^2 
% \end{eqnarray}
As the excitation becomes increasingly localized with randomness, the entanglement persists over longer periods of time (Fig.~\ref{fidelity}(a)-(b)), indicating that the time-evolved state retains the memory of the initial Bell state. This behavior is a \textit{clear signature of disorder-induced localization, which prevents quantum transport}.

\textit{Localization in higher-excitation subspaces:-} In excitation-conserving many-body models, the IPR with multiple excitations in the computational basis becomes a measure of localization in configuration space, not in real space~\cite{imbrie_ptamp,buijsman_scipost_2018}.  This measure includes correlations among the excitations and is, therefore, inadequate for the localization of individual particles.  Thus, in the two-excitation sector, two particles could be strongly localized in position, but their wavefunctions would remain spread out in the computational basis.
% Since we are interested in the localization of excitations in real space,  only IPR with one excitation sector can capture the overlap of a wavefunction with the real space, since configuration space and real space are identical in one excitation sector.

To bypass these IPR limitations,  we choose \textit{bipartite entanglement entropy} (EE) as a measure of ergodicity in the disordered system to quantify the entanglement between a subsystem and the rest of the system. In the ergodic phase, which satisfies the ETH~\cite{d2016quantum,deutsch2018eigenstate,von1929proof,reimann2010canonical,rigol2008thermalization,srednicki1994chaos,cramer2008exact}, the EE is expected to yield \textit{volume-law scaling}. By contrast, in the localized phase, it yields \textit{area-law scaling}~\cite{eisert_rmp_2010,philipp_prl_2017}. For $N$ TLS arranged in a one-dimensional array, we refer to the subsystem from $j=1$ to $j=N/2$ as $A$, so that its EE reads~\cite{roy_prb_2020,luitz_prb_2015,sur2025molecular,luitz_prb_2016} 
\begin{eqnarray}
S^{EE} = - Tr_A ~\rho_A \ln \rho_A,
\end{eqnarray}
where $\rho_A$ is the reduced density matrix of the subsystem $A$. This $S^{EE}$ scales linearly with $N$ in the ergodic phase, but in the localized phase $S^{EE}$  is generally independent of $N$. Since our model is excitation-conserving, we diagonalize the Hamiltonian for the two-excitation sector and compute the EE scaling~\cite{luitz_prb_2015}. %(Appendix C). 
% The entanglement entropy in the ergodic phase is expected to grow linearly with system size, although the proportionality constant depends on the excitation number and the maximum is obtained in the half-filled sector~\cite{garrison_prx_2018}.

{\color{black}
In a sector with \(n=\rho N\) excitations, where \(\rho\) is the excitation density or filling fraction, the global many-body level spacing decreases exponentially with system size as $\Delta_{\rm mb}\sim \frac{W}{\binom{N}{n}} \sim W e^{-N s(\rho)} $,
where \(s(\rho)\) is the entropy density. However, the stability of finite-density localization is not determined by this global many-body spacing alone~\cite{PhysRevLett.113.243002,PhysRevB.92.104428}, because a local spin-exchange process connects a given configuration only to a restricted subset of nearby configurations. For a power-law exchange interaction \(J(r)\sim J_0/r^\alpha\), the number of direct resonances in one dimension scales as
\begin{equation}
{\cal N}^{(1)}_{\rm res}\sim \frac{J_0}{W_{\rm loc}}\sum_{r=1}^{N}\frac{1}{r^\alpha}.
\end{equation}
Thus, direct long-distance resonances remain sparse for \(\alpha>1\), while the case \(\alpha=1\) is marginal (See SI). In the RDDI model, the near-zone coherent exchange behaves effectively as \(\alpha\simeq 3\), which favors finite-density localization, whereas the far-zone retarded interaction behaves as \(\alpha\simeq 1\), making it more susceptible to long-range resonances. A stronger finite-density instability can arise from resonant-pair hybridization; however, for two resonant pairs separated by a distance \(R\) much larger than their internal size \(a\), the leading long-range terms cancel, giving $K_{\rm pair}(R)\sim \frac{a^2}{R^{\alpha+2}}$. Therefore, in the near-zone dipolar regime with \(\alpha\simeq 3\), $K_{\rm pair}(R)\sim \frac{a^2}{R^5}$, which is summable at a large distance in one dimension. Hence, the effective pair-hybridization matrix elements decay faster than the bare microscopic exchange, supporting the stability of localization in the near-zone RDDI regime. This is consistent with the long-range \(XY\) resonance-counting criterion, according to which finite-density localization is stable for \(\alpha>3d/2\). For the present one-dimensional system, the near-zone exponent \(\alpha\simeq 3\) satisfies this condition, while the far-zone exponent \(\alpha\simeq 1\) does not.

}

{\color{black}The behavior of the eigenstate-averaged entanglement entropy $S^{EE}$, shown in Fig.~\ref{fig:ee}, as a function of system size $N$ in the $N/2$ and $N/4$ excitation sectors, clearly signals the transition from the delocalized to the localized phase. In the absence of disorder, $S^{EE}$ exhibits linear scaling with system size for sufficiently large $N$, irrespective of the TLS separation $R/\lambda_0$ (Fig.~\ref{fig:ee}(a)-(b)). This linear growth is characteristic of volume-law entanglement. In contrast, for strong positional disorder ($\delta R/R = 0.3$), the scaling behavior changes qualitatively. For $R/\lambda_0 = 0.05$ and $R/\lambda_0 = 0.2$, $S^{EE}$ no longer increases monotonically with $N$ (Fig.~\ref{fig:ee}(c)-(d)), and a clear deviation from linear scaling is observed. The suppression of entanglement growth with system size is consistent with the emergence of localization. Interestingly, for $R/\lambda_0 = 5.0$, linear growth of the averaged entanglement entropy is observed even in the presence of strong disorder. This suggests that when the inter-TLS separation is much larger than the characteristic wavelength, disorder becomes less effective in inducing localization.

}

% The behavior shown in Fig.~\ref{fig:ee} of $S^{EE}$, averaged over all eigenstates, as a function of the system size $N$ in the $N/2$ and $N/4$ excitation sector, clearly demarcates the transition from the delocalized to the localized phase. In the absence of disorder, $S^{EE}$ scales linearly with system size for sufficiently large $N$ irrespective of the TLS distance $R/\lambda_0$ (Fig.~\ref{fig:ee}(a) and (b)). In contrast, for high disorder strength $\delta R/R = 0.3$, $S^{EE}$ does not increase monotonically (Fig.~\ref{fig:ee}(c) and (d)) for $R/\lambda_0 = 0.05$ and $R/\lambda_0 = 0.2$. Interestingly, the linear growth of averaged EE is observed for $R/\lambda_0 = 5.0$. This result indicates a strong localization in the system when the inter-TLS distance is either of the characteristic wavelength or smaller than that. 

\textit{Discussion:-} 
The occurrence of localization via long-range randomized couplings in our model suggests a critical role of system-specific interactions, such as RDDI, \textit{which do not conform to scalar power-law models}~\cite{abanin_2019_rmp,basko2006metal,gornyi2005interacting,alet2018many,abanin2019colloquium}. The vectorial nature of the long-range photon-mediated (retarded) dipolar interaction underlies the resulting unfamiliar MBL behavior, particularly its resilience to decay and dissipation. These features make the RDDI-coupled position-disordered $N-$TLS system an intriguing platform for exploring localization transitions in open quantum systems, thus extending the scope of known localization physics.

Our findings open a novel route to understanding off-diagonal disorder-driven confinement. These localized modes induced by RDDI randomness exhibit sharp spatial confinement and \textit{insensitivity to system size and disorder strength}, thus providing a promising mechanism for robust quantum memory. On the applied side, we have shown that off-diagonal disorder in RDDI-coupled systems leads to excitation confinement that is robust to dissipation, which is inherent in RDDI. Demonstration of the fidelity preservation of Bell states further emphasizes the importance of our model for potential applications in quantum technologies.
The pursuit of long-lived, robust, and scalable quantum memories~\cite{chaneliere2005storage,hedges2010efficient,appel2008quantum,hashimoto2019all,li2020controlled,moiseev2025optical,jing2023ensemble,wang2025entanglement,zhu2025remote,liu2025millisecond,rui2015operating,julsgaard2013quantum} is commonly confronted by thermalization and environmental decoherence.

Our model may elucidate the physics of exciton transport in biological complexes like the Fenna–Matthews–Olson (FMO) system, where energy is funneled across pigment-protein networks via dipole-mediated excitonic hopping~\cite{engel2007evidence,briggs2011equivalence}, while disorder, whether structural or energetic, can induce localization and suppress transport~\cite{rey2013exploiting}. 
Similar physics governs exciton transport in semiconductor quantum dots and molecular aggregates dominated by Förster-type dipolar interactions~\cite{akselrod2014subdiffusive,romero2014quantum}. Variations in site spacing, RDDI coupling, and disorder strength lead to a crossover from ballistic to localized transport, captured by our results~\cite{creatore2013efficient,feierabend2017proposal}.

% Similar physics governs exciton motion in semiconductor quantum dots and molecular aggregates, where F\"orster-type dipolar interactions dominate~\cite{akselrod2014subdiffusive,romero2014quantum}. Here, tuning site spacing, RDDI coupling, and disorder strength induce a crossover from ballistic to localized transport, which our results capture. These insights may contribute to bio-inspired quantum technologies, including coherent photovoltaics~\cite{creatore2013efficient}, excitonic circuits, and quantum sensors~\cite{feierabend2017proposal}. 

% To conclude, our results indicate that by tailoring structural disorder and RDDI geometry, it may be possible to guide, trap, or selectively route excitations in the quantum domain~\cite {rey2013exploiting}.

\textit{Acknowledgement: } P.C. acknowledges support from the International Postdoctoral Fellowship from the Ben May Center for Theory and Computation. A.M. acknowledges support from the Anusandhan National Research Foundation, Government of India (Grant No. ANRF/ECRG/2024/003836/PMS). D.P. was supported by the EU HORIZON-RIA Project EuRyQa (Grant No. 101070144) and the HESC of Armenia (Grants No. 24FP-1F03). A.G. acknowledges support from the Anusandhan National Research Foundation, Government of India (Grant No.  ANRF/ARGM/2025/000622/TS).

\bibliography{lit-therm-coh}

 \onecolumngrid
% \appendix

% \newpage

 \appendix

\begin{center}
    \Large End Matter
\end{center}

\section{Scaling of IPR}
\label{app:A1}

\begin{figure}[h]
  \begin{center}
  \includegraphics[width=0.76\textwidth]{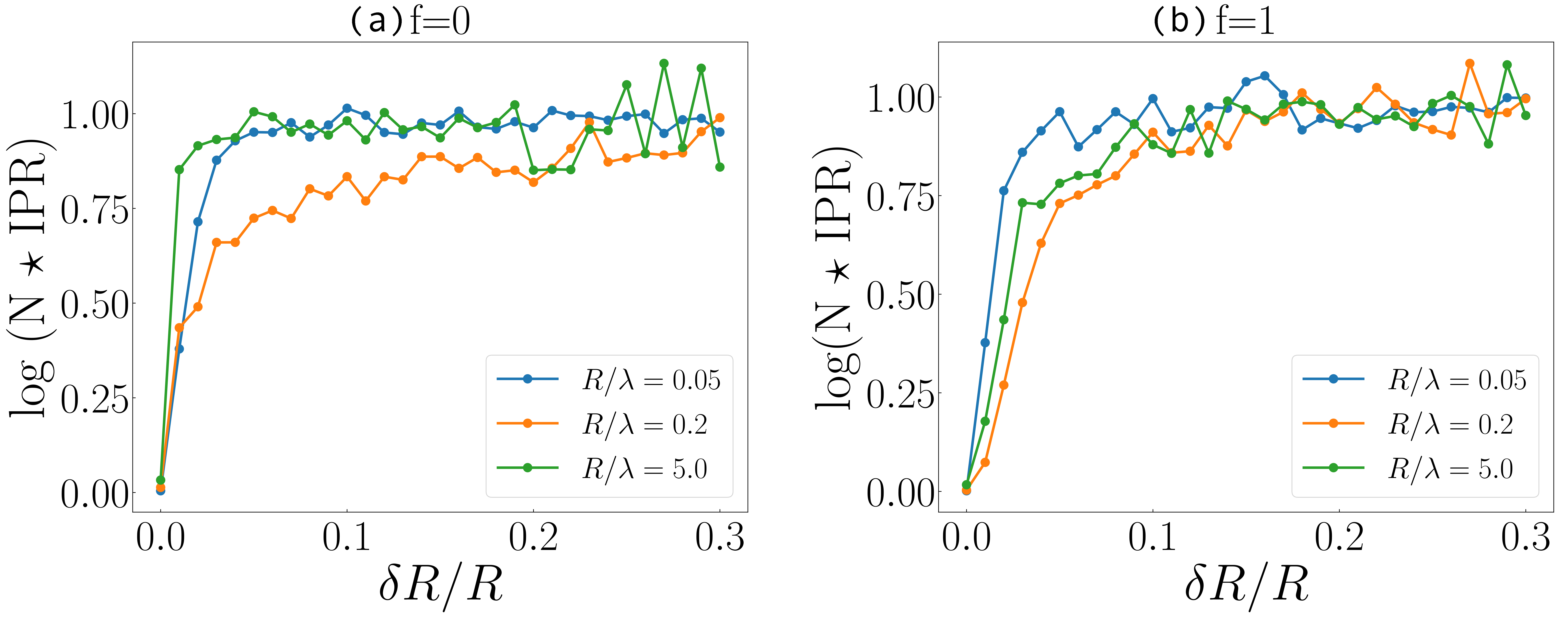}
  \end{center}
      \caption{The variation of the  $\log (N*IPR)$ with the system size $\log N$ with the off-diagonal disorder indicates the transition from the extended state to the localized state for (a) $f=0$, dispersive RDDI, where switching on the randomness results in the transition, and for (b) $f=1$, dissipative RDDI, where the transition is more smooth than the case without the dissipation. }
      \label{scaling_WdisM}
  \end{figure}

The IPR exhibits a characteristic scaling with the system size, described by the relation $1/N^\delta$, with  $\delta =1$ for a conventional extended state in a one-dimensional network. For the extended states, $\log (N*IPR)$ remains invariant with increasing system size $N$, reflecting a uniform distribution of the eigenstate across the entire system. In contrast, for localized states,  $\log (N*IPR)$ will increase with the $\log N$, indicating that the effective spatial extent of the eigenstate remains finite as the system grows. This distinction provides a robust criterion to differentiate between extended and localized phases, and the transition between them is clearly visualized in Fig.~\ref{scaling_WdisM}.  The scaling of the IPR with the system size for the short and long-range interactions with and without dissipation for different disorder strengths is shown in Fig.~\ref{scaling_WdisM}.

The figure presents the scaling behavior of the IPR across systems with short-range and long-range interactions, both in the presence and absence of dissipation, for varying strengths of positional disorder. The scaling analysis unequivocally demonstrates a transition from extended to localized states as disorder is introduced, regardless of the interaction range or the presence of dissipation. Notably, even under long-range interactions, typically considered to support delocalized eigenstates, the introduction of sufficient disorder induces localization. This is a remarkable departure from conventional wisdom in disordered one-dimensional systems with power-law couplings, where it is generally accepted that no localization occurs for interaction exponents $\alpha < 2$~\cite{levitov1989absence,mirlin2000statistics}.

 \begin{figure}
    \centering
 \includegraphics[width=0.76\textwidth]{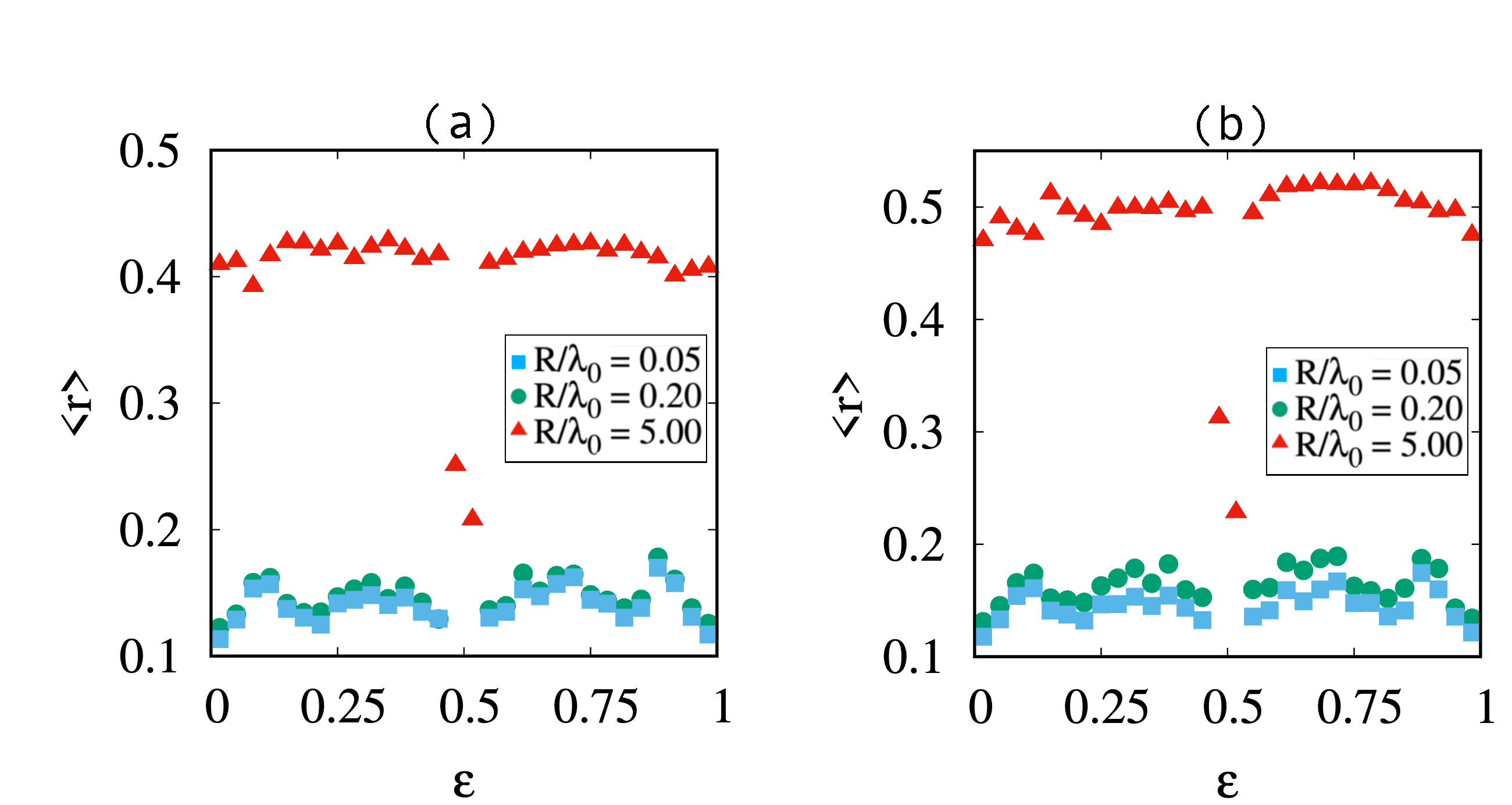}
    \caption{ {Energy-resolved average adjacent gap ratio $\langle r \rangle$ as a function of normalized energy density $\epsilon$ for different interaction wavelengths $R/\lambda_0$ in the half-filled sector of the disordered RDDI model for (a) $f=0$ and (b) $f=1$. The
gap ratio is computed from the spacings of the real parts of the complex
eigenvalues for $f=1$.  The system size is $N=14$ with disorder $\delta R /R = 0.3$ for this plot.} }
    \label{mobility_edge}
\end{figure}

{\color{black} The term $V_{ij}$ represents the coherent part of the resonant dipole-dipole interaction between two emitters. This term describes the exchange of virtual photons, leading to coherent excitation transfer and effective long-range hopping in the system. In contrast, the term $F_{ij}$ characterizes collective radiative dissipation arising from coupling to the common electromagnetic environment. It accounts for cooperative decay processes and contributes to the anti-Hermitian part of the effective open-system dynamics. Both $V_{ij}$ and $F_{ij}$ originate from the same electromagnetic Green’s function and therefore possess similar spatial dependence on the inter-emitter separation $R_{ij}$, differing mainly through sine and cosine contributions~\cite{stephen_jcp_1964}.

Our numerical results indicate that the inclusion of the dissipative term $F_{ij}$ does not qualitatively alter the localization behavior of the system. This can be attributed to the fact that both $V_{ij}$ and $F_{ij}$ exhibit similar oscillatory and algebraically decaying dependence on $R_{ij}$. As a result, the dissipative coupling modifies the spectrum quantitatively while leaving the essential localization properties largely unchanged.}

\color{black}
\section{Mobility edge}

The energy-resolved mean gap-ratio analysis in Fig.~\ref{mobility_edge} does not show any significant variation of $\langle r \rangle$ across the many-body spectrum for a fixed interaction wavelength for both with and without dissipation. This indicates the absence of a clear many-body mobility edge in the present system. Instead, we see that the spectral statistics remain approximately uniform throughout the spectrum, with the nature of the statistics being predominantly controlled by the interaction wavelength $R/\lambda_0$. Particularly, for $R/\lambda_0 = 0.05$, the system exhibits weak level repulsion with $\langle r \rangle \approx 0.42$, suggestive of a critical or nonergodic extended regime, whereas for larger wavelengths the significantly suppressed values of $\langle r \rangle$ indicate anomalous nonergodic behavior. A plausible origin of the absence of a mobility edge is the long-range oscillatory nature of the RDDI couplings, which globally reorganize the resonance structure of the many-body Hilbert space rather than producing energy-selective localization transitions.

% % 
\clearpage
\input{Supplementary_arxiv.tex}

\end{document}

%% file: Supplementary_arxiv.tex
 %auto-ignore

%\usepackage{epsfig,graphicx}
\linespread{1.0}

\begin{center}
    \textbf{\large Supplementary Material: Quantum state localization in dipole-dipole interacting disordered networks}
\end{center}

\author{Pritam Chattopadhyay}
\thanks{contributed equally}
% \email{pritam.chattopadhyay@weizmann.ac.il}
\affiliation{Department of Chemical and Biological Physics \& AMOS,
Weizmann Institute of Science, Rehovot 7610001, Israel}

\author{Saikat Sur}
\thanks{contributed equally}
% \email{saikat.sur@weizmann.ac.il}
\affiliation{Department of Chemical and Biological Physics \& AMOS,
Weizmann Institute of Science, Rehovot 7610001, Israel}
\affiliation{Optics \& Quantum Information Group, The Institute of Mathematical Sciences, HBNI, CIT Campus, Taramani, Chennai 600113, India}

\author{Avijit Misra}
% \thanks{contributed equally}
% \email{avijitmisra@iitism.ac.in}
% \affiliation{Department of Chemical and Biological Physics \& AMOS,
% Weizmann Institute of Science, Rehovot 7610001, Israel}
 \affiliation{Department of Physics, Indian Institute of Technology (ISM), Dhanbad, Jharkhand 826004, India}

\author{David Petrosyan}
% \email{dap@iesl.forth.gr}
\affiliation{Institute of Electronic Structure and Laser, FORTH, GR-70013 Heraklion, Greece}

\author{Arti Garg}
% \email{arti.garg@saha.ac.in}
\affiliation{Theory Division, Saha Institute of Nuclear Physics, 1/AF Bidhannagar, Kolkata 700 064, India}
\affiliation{Homi Bhabha National Institute, Training School Complex, Anushaktinagar, Mumbai 400094, India}

\author{Gershon Kurizki}
% \email{gershon.kurizki@weizmann.ac.il}
\affiliation{Department of Chemical and Biological Physics \& AMOS,
Weizmann Institute of Science, Rehovot 7610001, Israel}

\maketitle

\onecolumngrid

\color{black}
\section{ Jordan-Wigner transformation of RDDI model }\label{AppA}

The Jordan-Wigner transformation~\cite{jordan_zfp_1928} from spin operators to fermion creation/annihilation operators is given as
\begin{subequations}
\begin{eqnarray}
  \sigma^+_j = c^\dagger_j e^{i\pi \sum_{k<j} c^\dagger_k c_k},    
\end{eqnarray}
\begin{eqnarray}
  \sigma^-_j =  e^{-i\pi \sum_{k<j} c^\dagger_k c_k}c_j,    
\end{eqnarray}
\begin{eqnarray}
  \sigma^z_j =  2c^\dagger_j c_j - \mathbf{1}.    
\end{eqnarray}
\end{subequations}

The Hamiltonian in terms of fermion creation and annihilation operators reads 

\begin{eqnarray}
  H &=&  2\sum^N_{j=1} \omega_0 c^\dagger_j c_j + \sum^{N}_{j,j'=1; j<j'} M_{jj'} (\theta,R_{jj'}) \left(c^\dagger_j \prod^{j'-1}_{k=1} (\mathbf{1} - 2 c^\dagger_k c_k) c_{j'} + \text{h.c.}\right)\nonumber\\
  &=& 2\sum^N_{j=1} \omega_0 c^\dagger_j c_j + \sum^{N}_{j,j'=1; j<j'} M_{jj'} (\theta,R_{jj'}) \Big(c^\dagger_j c_{j'} - 2 \sum_{j \le k \le (j'-1)} c^\dagger_j c^\dagger_k c_k c_{j'}  + 4 \sum_{j \le k \le k'\le (j'-1)}   c^\dagger_j c^\dagger_k c_k c^\dagger_{k'} c_{k'} c_{j'} + ... + \text{h.c.}\Big).
\end{eqnarray}

For long-range RDDI couplings beyond nearest neighbor in one dimension, the Jordan-Wigner string $\prod^{j'-1}_{k=1} (\mathbf{1} - 2 c^\dagger_k c_k)$  survives. As a result, we obtain nonlocal terms with all possible even numbers of fermion operators. Thus, the  RDDI Hamiltonian consists of an infinite sum of such terms in the fermion language. Only for nearest neighbor couplings, the Hamiltonian becomes bilinear in terms of Fermionic creation and annihilation operators. 

\color{black}

\color{black}
% \begin{figure}
%     \centering
%     \includegraphics[width=0.98\linewidth]{prob_density_pertb_av_2.pdf}
%     \caption{Caption}
%     \label{fig_pertb1}
% \end{figure}

\section{Perturbative Treatment of Disorder-Induced Localization}\label{App B}

\begin{figure}[t]
    \centering
 \includegraphics[width=0.75\textwidth]{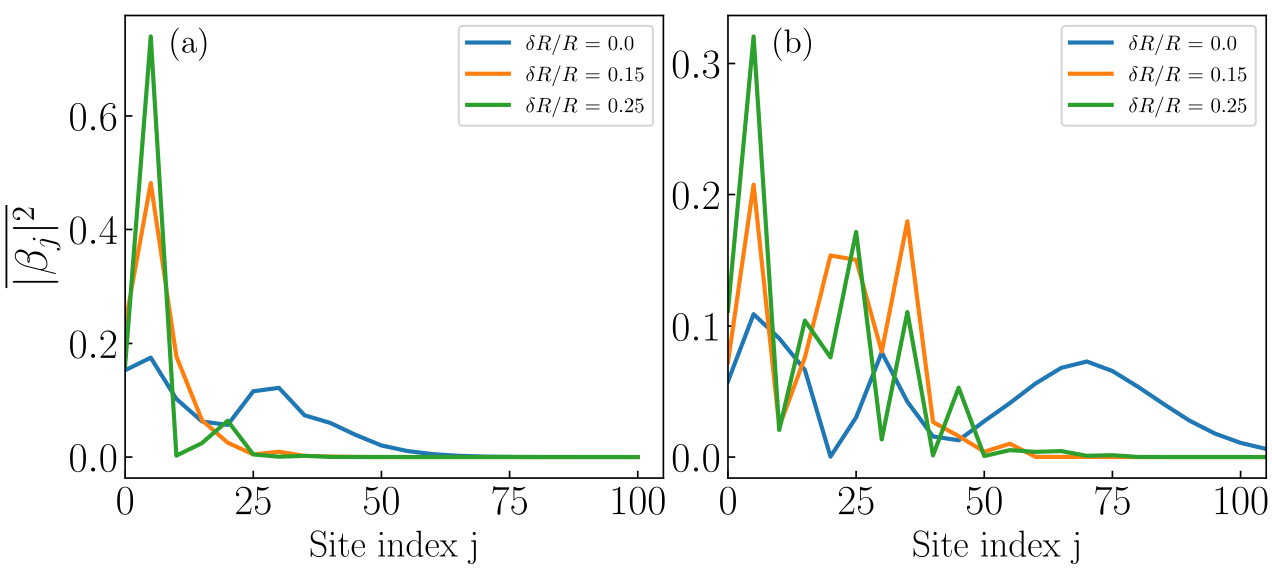}
 \caption{ The overlap of the single-atom states $\vert e_j\rangle$ with the perturbed single excitation eigenvector $\vert \phi^{(1)}_k \rangle$  averaged over all eigenvectors: $ \overline{{\vert \beta_j\vert^2}} = \frac{1}{N} \sum_k \vert \langle e_j|\phi_k^{(1)}\rangle|^2$ with $N=100$ for interatomic distances (a) $R/\lambda_0=5.00$, (b) $R/\lambda_0=0.20$, %and (c) $R/\lambda_0=0.05$ 
     for different disorder strengths. In the clean case, the excitation is distributed almost uniformly across the system, which is characteristic of extended states. As disorder increases, the excitation weight becomes concentrated over a few sites, signaling spatial confinement of the excitation. Such confinement with increasing disorder provides direct evidence of localization. The plots are generated after averaging the data from twenty independent disorder configurations from eigenstates computed from the first-order perturbation in disorder strength.}
    \label{perturb}
\end{figure}

The RDDI terms in the Hamiltonian (Eq. (1a) of the main text) change in the presence of positional disorder:
\begin{subequations}
\begin{equation}
    M_{jj'}^{(1)}(\theta,R_{jj'}) = M_{jj'}^{(0)}(\theta,R^0_{jj'}) +  \frac{dM(\theta,R_{jj'})}{dR}\Big|_{R = R_{jj'}^0} \delta R_{jj'} + \mathcal{O}(\delta R^2),
\end{equation}
with $\delta R_{jj'} = \delta X_j - \delta X_{j'}$, and entail changes in the Hamiltonian 
\begin{equation}
    H\rightarrow H+H'.
\end{equation}
\end{subequations}
To assess the disorder effects, we expand the perturbed $k$th eigenfunction to first-order in the disorder-induced  (complex) random correction as %$H'_{jj'}$ as with variance $\sim \Delta_X^2$ as 
\begin{equation} \label{eqn3}
    |\phi_k^{(1)}\rangle = | \phi_k^{(0)} \rangle + \sum_{l \ne k} \frac{\langle \phi_l^{(0)} | H' | \phi_k^{(0)} \rangle}{E_k^{(0)} - E_l^{(0)}} |\phi_l^{(0)}\rangle.
\end{equation}
which is a sum over many complex terms reflecting the randomness in the first-order perturbation $H'$ in the translationally invariant Hamiltonian.
% This sum of the many extended states with random phases introduces random fluctuations that disrupt the spatial \textit{uniformity} of the extended unperturbed eigenstates, leading to their strong spatial variations. Further insight may be gained by recalling that the 
Each unperturbed zeroth-order eigenfunction $|\phi_k^{(0)}\rangle$ can be interpreted as a coherent sum over \textit{multiple-scattering interfering trajectories} between sites~\cite{Akkermans2007,lehmberg_pra_1970}. In a translationally invariant system, this sum is delocalized, but in the presence of disorder, which renders the path-dependent phases random, the incoherent summation in \eqref{eqn3} replaces the extended waves $|\phi_k^{(0)}\rangle$ by strongly fluctuating perturbed eigenfunctions $|\phi_k^{(1)}\rangle$. The wavefunction amplitude becomes \textit{statistically larger in a small cluster of sites while being exponentially suppressed elsewhere} (Fig. \ref{perturb}).

\subsection{First-order perturbative expansion of the dispersive (real) kernel of RDDI}
We consider the coherent (dispersive) dipole-dipole interaction term between sites 
$j_1$ and $j_2$, given by
\begin{equation}
V_{j_1 j_2}(\theta, R_{j_1 j_2}) 
= \frac{\gamma}{4} \Bigg[ (1 - 3 \cos^2\theta) 
\left( \frac{\cos(k_0 R_{j_1 j_2})}{(k_0 R_{j_1 j_2})^3} 
+ \frac{\sin(k_0 R_{j_1 j_2})}{(k_0 R_{j_1 j_2})^2} \right) 
- (1 - \cos^2\theta) \frac{\cos(k_0 R_{j_1 j_2})}{k_0 R_{j_1 j_2}} \Bigg].
\label{eq:Vkernel}
\end{equation}

%\subsection*{Disorder in site positions}

Let the equilibrium (clean lattice) positions be $X_j^0$, with small displacements 
$\delta X_j$ due to positional disorder. The intersite distance becomes
\begin{equation}
R_{j_1 j_2} = R_{j_1 j_2}^0 + \delta R_{j_1 j_2}, 
\qquad 
\delta R_{j_1 j_2} = \delta X_{j_1} - \delta X_{j_2}.
\end{equation}

Here $R_{j_1 j_2}^0 = |X_{j_1}^0 - X_{j_2}^0|$ is the separation.
We expand the kernel to first order in the displacement $\delta R_{j_1 j_2}$
\begin{equation}
V_{j_1 j_2}(R_{j_1 j_2}) 
= V_{j_1 j_2}(R_{j_1 j_2}^0 + \delta R_{j_1 j_2})
\simeq V_{j_1 j_2}^0 
+ V_{j_1 j_2}'(R_{j_1 j_2}^0)\, \delta R_{j_1 j_2},
\end{equation}
where
\begin{equation}
V_{j_1 j_2}^0 := V_{j_1 j_2}(R_{j_1 j_2}^0),
\qquad 
V_{j_1 j_2}'(R) := \frac{d}{dR} V_{j_1 j_2}(R).
\end{equation}

Thus, the linear perturbation reads
\begin{equation}
\delta V_{j_1 j_2} \simeq V_{j_1 j_2}'(R_{j_1 j_2}^0) \, (\delta X_{j_1} - \delta X_{j_2}).
\end{equation}

%\subsection*{Derivative of the kernel}

From Eq.~\eqref{eq:Vkernel}, we set $x = k_0 R$ and compute
\begin{align}
V(R) &= \frac{\gamma}{4} \left[ (1-3\cos^2\theta) 
\left( \frac{\cos x}{x^3} + \frac{\sin x}{x^2} \right) 
- (1-\cos^2\theta) \frac{\cos x}{x} \right],
\end{align}
where $x = k_0 R$. Differentiating with respect to $R$
\begin{align}
\frac{dV}{dR} &= \frac{\gamma}{4} \left[ (1-3\cos^2\theta) \frac{d}{dR}\left( \frac{\cos x}{x^3} + \frac{\sin x}{x^2} \right) 
- (1-\cos^2\theta) \frac{d}{dR}\left(\frac{\cos x}{x}\right) \right].
\end{align}

Using the chain rule $\frac{d}{dR} = k_0 \frac{d}{dx}$, we compute term by term:
\begin{align}
\frac{d}{dx}\left(\frac{\cos x}{x^3}\right) &= -\frac{\sin x}{x^3} - \frac{3\cos x}{x^4}, \\
\frac{d}{dx}\left(\frac{\sin x}{x^2}\right) &= \frac{\cos x}{x^2} - \frac{2\sin x}{x^3}, \\
\frac{d}{dx}\left(\frac{\cos x}{x}\right) &= -\frac{\sin x}{x} - \frac{\cos x}{x^2}.
\end{align}

Therefore,
\begin{align}
\frac{dV}{dR} = \frac{\gamma k_0}{4} \Bigg[ (1-3\cos^2\theta) 
\left( -\frac{\sin x}{x^3} - \frac{3\cos x}{x^4} + \frac{\cos x}{x^2} - \frac{2\sin x}{x^3} \right)  - (1-\cos^2\theta)\left( -\frac{\sin x}{x} - \frac{\cos x}{x^2} \right) \Bigg].
\end{align}

%\subsection*{Final linear perturbation}

Thus, the first-order correction to the matrix element is
\begin{equation}
\delta V_{j_1 j_2} 
\simeq \frac{dV}{dR}\bigg|_{R=R_{j_1 j_2}^0} \, (\delta X_{j_1} - \delta X_{j_2}),
\end{equation}
where $\tfrac{dV}{dR}$ is explicitly given above. Collecting the terms yields the closed form
\begin{equation}
V'(R)=\frac{\gamma k_0}{4}\Bigg[
\frac{B\,\sin(k_0R)}{k_0R} \;-\; \frac{3A\,\sin(k_0R)}{(k_0R)^3}
+ \cos(k_0R)\!\left(\frac{A+B}{(k_0R)^2} - \frac{3A}{(k_0R)^4}\right)
\Bigg].
\label{eq:Vprime_closed}
\end{equation}

\subsection{First-order perturbative expansion of the dissipative (imaginary) kernel of RDDI }

We complement the dispersive analysis with the imaginary (radiative/dissipative) RDDI component.  
The dissipative kernel between atoms $j_1$ and $j_2$ is given by
\begin{equation}
F_{j_1 j_2}(\theta,R_{j_1 j_2})
= \frac{f\gamma}{4}\Bigg[
(-1+3\cos^2\theta)\Big(\frac{\sin(k_0R_{j_1 j_2})}{(k_0R_{j_1 j_2})^3}
-\frac{\cos(k_0R_{j_1 j_2})}{(k_0R_{j_1 j_2})^2}\Big)
+ (1-\cos^2\theta)\frac{\sin(k_0R_{j_1 j_2})}{k_0R_{j_1 j_2}}
\Bigg].
\label{eq:Fkernel}
\end{equation}

%\subsection*{Positional disorder and linearization}

As in the dispersive case, we introduce small positional disorder,
\[
X_j \mapsto X_j + \delta X_j,\qquad 
\bar{\delta X_j}=0,\quad \mathrm{Var}(\delta X_j)=\Delta_X^2,\quad \Delta_X\ll a,
\]
and define the signed change in separation
\[
\delta R_{j_1 j_2}=\delta X_{j_1}-\delta X_{j_2}.
\]
To first order in $\delta R$, we expand
\[
F(R_{j_1 j_2}^0+\delta R_{j_1 j_2})
\simeq F(R_{j_1 j_2}^0) + F'(R_{j_1 j_2}^0)\,\delta R_{j_1 j_2},
\]
where $F'(R)=\dfrac{dF}{dR}(R)$.

Thus, the linear-order imaginary perturbation in the site basis is (for $j_1\neq j_2$)
\begin{equation}
H'^{(F)}_{j_1 j_2} \approx F'(R^0_{j_1 j_2})\,(\delta X_{j_1}-\delta X_{j_2}).
\end{equation}

%\subsection*{Derivative $F'(R)$ in closed form (use $x=k_0R$)}

Set $x\equiv k_0R,\qquad C\equiv -1+3\cos^2\theta,\qquad D\equiv 1-\cos^2\theta$,
so that the kernel \eqref{eq:Fkernel} reads
\begin{equation}
F(R) = \frac{f\gamma}{4}\Big[ C\Big(\frac{\sin x}{x^3} - \frac{\cos x}{x^2}\Big) + D\frac{\sin x}{x}\Big].
\end{equation}

Differentiate with respect to $R$ using $\dfrac{d}{dR}=k_0\dfrac{d}{dx}$. Compute termwise derivatives with respect to $x$
\begin{align}
\frac{d}{dx}\!\Big(\frac{\sin x}{x^3}\Big) &= \frac{\cos x}{x^3} - \frac{3\sin x}{x^4},\\
\frac{d}{dx}\!\Big(-\frac{\cos x}{x^2}\Big) &= \frac{\sin x}{x^2} + \frac{2\cos x}{x^3},\\
\frac{d}{dx}\!\Big(\frac{\sin x}{x}\Big) &= \frac{\cos x}{x} - \frac{\sin x}{x^2}.
\end{align}

Collecting these contributions gives
\begin{align}
\frac{dF}{dx}
&= \frac{f\gamma}{4}\Bigg[
C\Big(\frac{\cos x}{x^3} - \frac{3\sin x}{x^4} + \frac{\sin x}{x^2} + \frac{2\cos x}{x^3}\Big)
+ D\Big(\frac{\cos x}{x} - \frac{\sin x}{x^2}\Big)\Bigg]\nonumber\\[4pt]
&= \frac{f\gamma}{4}\Bigg[
D\frac{\cos x}{x} - 3C\frac{\sin x}{x^4} + \sin x\Big(\frac{C}{x^2} - \frac{D}{x^2}\Big)
+ \cos x\Big(\frac{3C}{x^3} + \frac{C}{x^3}\Big)\Bigg],
\end{align}
which can be simplified and rearranged to a numerically convenient form. Applying the chain rule,
\begin{equation}
F'(R)=\frac{f\gamma k_0}{4}\Bigg[
\frac{D\cos(k_0R)}{k_0R} - \frac{3C\sin(k_0R)}{(k_0R)^4}
+ \frac{(C-D)\sin(k_0R)}{(k_0R)^2}
+ \frac{4C\cos(k_0R)}{(k_0R)^3}
\Bigg].
\label{eq:Fprime}
\end{equation}

%(You may rearrange algebraically to optimize numerical stability; the displayed form above exhibits the different algebraic decays.)

% \subsection*{Asymptotic behaviour of $F'(R)$}

% \paragraph{Short-distance} ($k_0R \ll 1$): use $\sin x\approx x - x^3/6$ and $\cos x\approx 1 - x^2/2$. The most singular contribution comes from the $\sin x / x^4$ term and one finds
% \[
% F'(R)\underset{x\ll1}{\sim} -\,\frac{3f\gamma C}{4}\,\frac{1}{k_0^3 R^4}\;\propto\; -\frac{1}{R^4},
% \]
% i.e. the same $R^{-4}$ scaling as the dispersive derivative (up to angular prefactors).

% \paragraph{Large-distance} ($k_0R\gg1$): the leading term decays as $\cos(k_0R)/R$ or $\sin(k_0R)/R^2$ depending on prefactors; the envelope is algebraic:
% \[
% F'(R)\underset{x\gg1}{\sim} \frac{f\gamma}{4}\,D\,\frac{\cos(k_0R)}{R}.
% \]

% \subsection*{Combined complex derivative and linear perturbation}

Define the complex derivative of the full kernel
$M'(R)= \frac{d}{dR}M(R) = V'(R) + iF'(R)$
Then, in linear order, the full perturbation matrix element in the site basis is
\begin{equation}
H'_{j_1 j_2} \approx M'(R^0_{j_1 j_2})\,(\delta X_{j_1}-\delta X_{j_2}) = \big(V'(R^0_{j_1 j_2}) + i F'(R^0_{j_1 j_2})\big)\,(\delta X_{j_1}-\delta X_{j_2}).    \end{equation}

Because the $\delta X_j$ are independent and $\mathrm{Var}(\delta X_j-\delta X_{j'})=2\Delta_X^2$, we find
\[
\mathrm{Var}\big(H'_{j_1 j_2}\big) = 2\Delta_X^2\,|M'(R^0_{j_1 j_2})|^2
= 2\Delta_X^2\Big([V'(R^0_{j_1 j_2})]^2 + [F'(R^0_{j_1 j_2})]^2\Big).
\]

\section{ Off-diagonal disorder }
\label{app:C}
\subsection{Scaling of IPR}

\begin{figure*}
  \begin{center}
\includegraphics[width=0.85\textwidth]{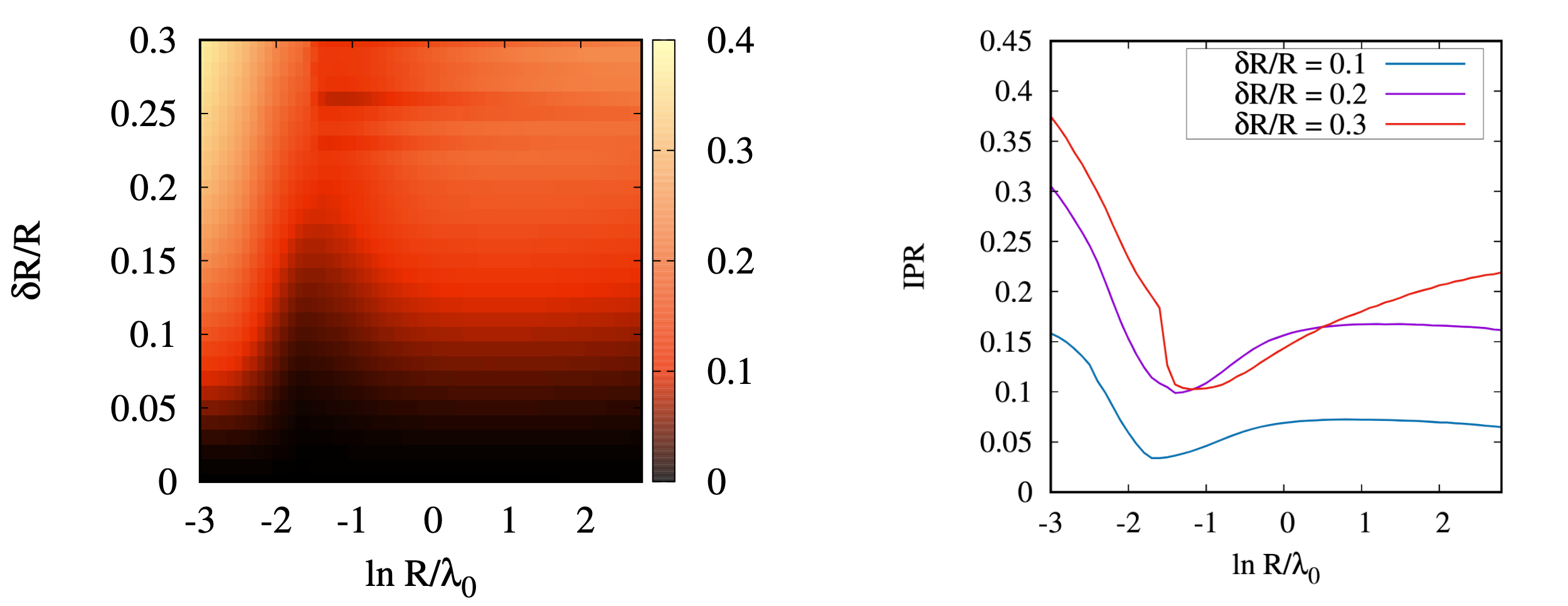}
\end{center}
      \caption{The static inverse participation ratio (IPR) for the off-diagonal disorder averaged over all eigenvectors with $\ln(R/\lambda_0)$ for $N = 1000$ with dissipation for $f=1$.}
      \label{ColorPlot_WdisM1}
  \end{figure*}

\begin{figure*}
  \begin{center}
  \subfigure[]{%
  \includegraphics[width=0.48\textwidth]{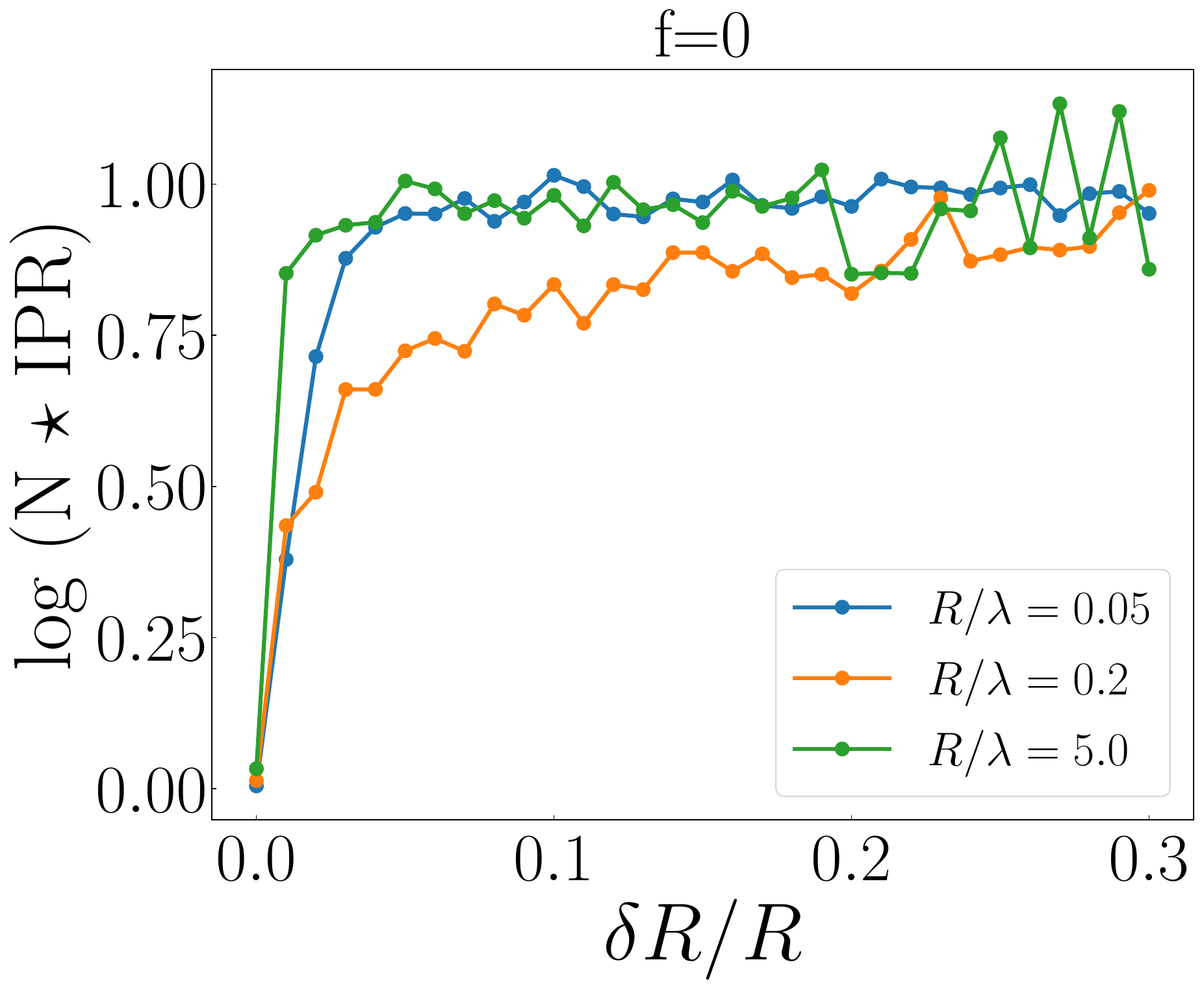}}
         \subfigure[]{%
  \includegraphics[width=0.48\textwidth]{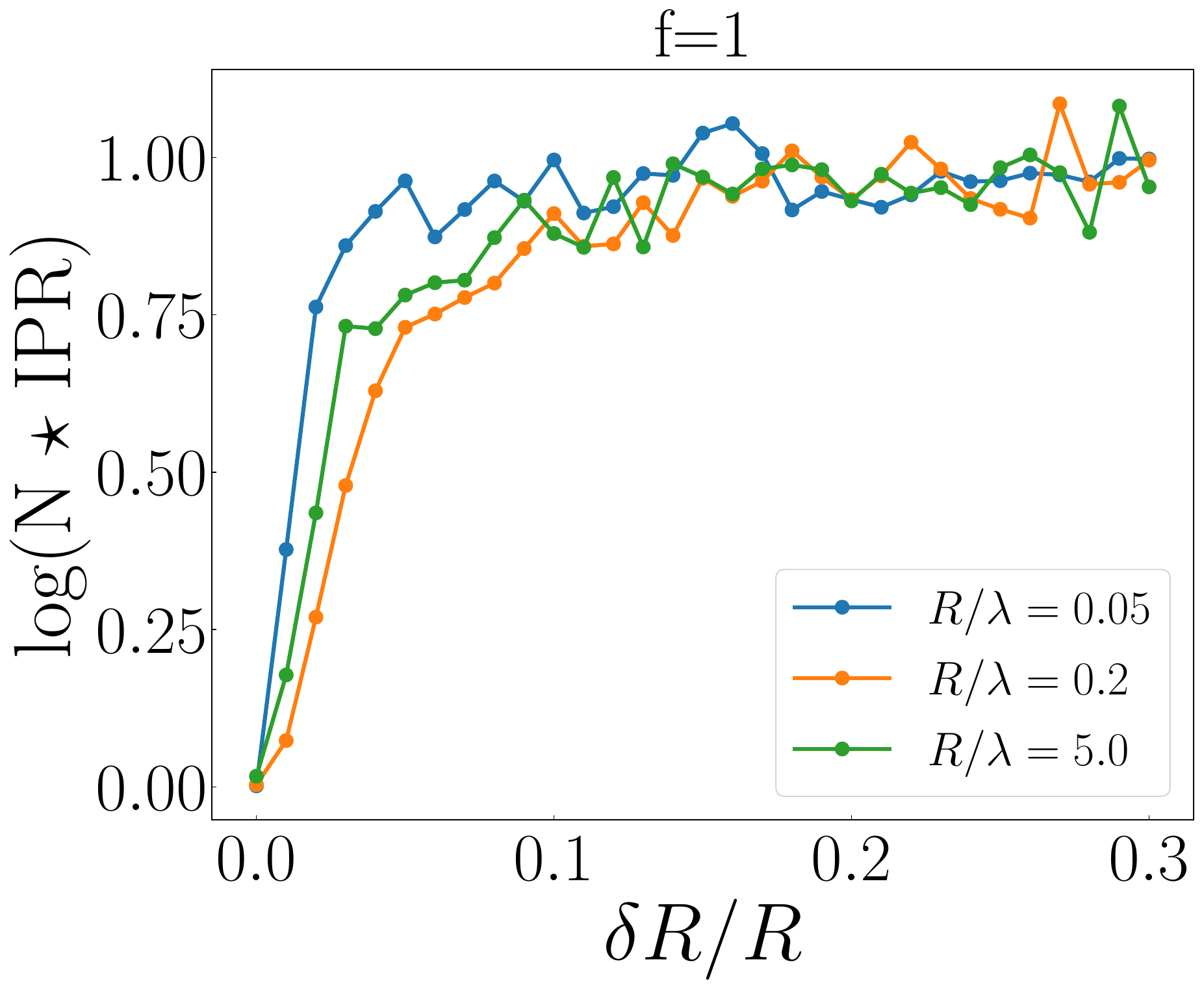}}
  \end{center}
      \caption{The variation of the slope (derived from the variation of $\log (N*IPR)$ with the system size $\log N$) with the off-diagonal disorder indicates the transition from the extended state to the localized state for (a) $f=0$, where switching on the randomness results in the transition, and for (b) $f=1$, where the transition is more smooth than the case without the dissipation. }
      \label{scaling_WdisM}
  \end{figure*}

The IPR exhibits a characteristic scaling with the system size, described by the relation $1/N^\delta$, with  $\delta =1$ for a conventional extended state in a one-dimensional network. For the extended states, $\log (N*IPR)$ remains invariant with increasing system size $N$, reflecting a uniform distribution of the eigenstate across the entire system. In contrast, for localized states,  $\log (N*IPR)$ will increase with the $\log N$, indicating that the effective spatial extent of the eigenstate remains finite as the system grows. This distinction provides a robust criterion to differentiate between extended and localized phases, and the transition between them is clearly visualized in Fig.~\ref{scaling_WdisM}.  The scaling of the IPR with the system size for the short and long-range interactions with and without dissipation for different disorder strengths is shown in Fig.~\ref{scaling_WdisM}. 

The figure presents the scaling behavior of the IPR across systems with short-range and long-range interactions, both in the presence and absence of dissipation, for varying strengths of positional disorder. The scaling analysis unequivocally demonstrates a transition from extended to localized states as disorder is introduced, regardless of the interaction range or the presence of dissipation. Notably, even under long-range interactions, typically considered to support delocalized eigenstates, the introduction of sufficient disorder induces localization. This is a remarkable departure from conventional wisdom in disordered one-dimensional systems with power-law couplings, where it is generally accepted that no localization occurs for interaction exponents $\alpha < 2$~\cite{levitov1989absence,mirlin2000statistics}.

\begin{figure*}
  \begin{center}
       \subfigure[]{%
  \includegraphics[width=0.45\textwidth]{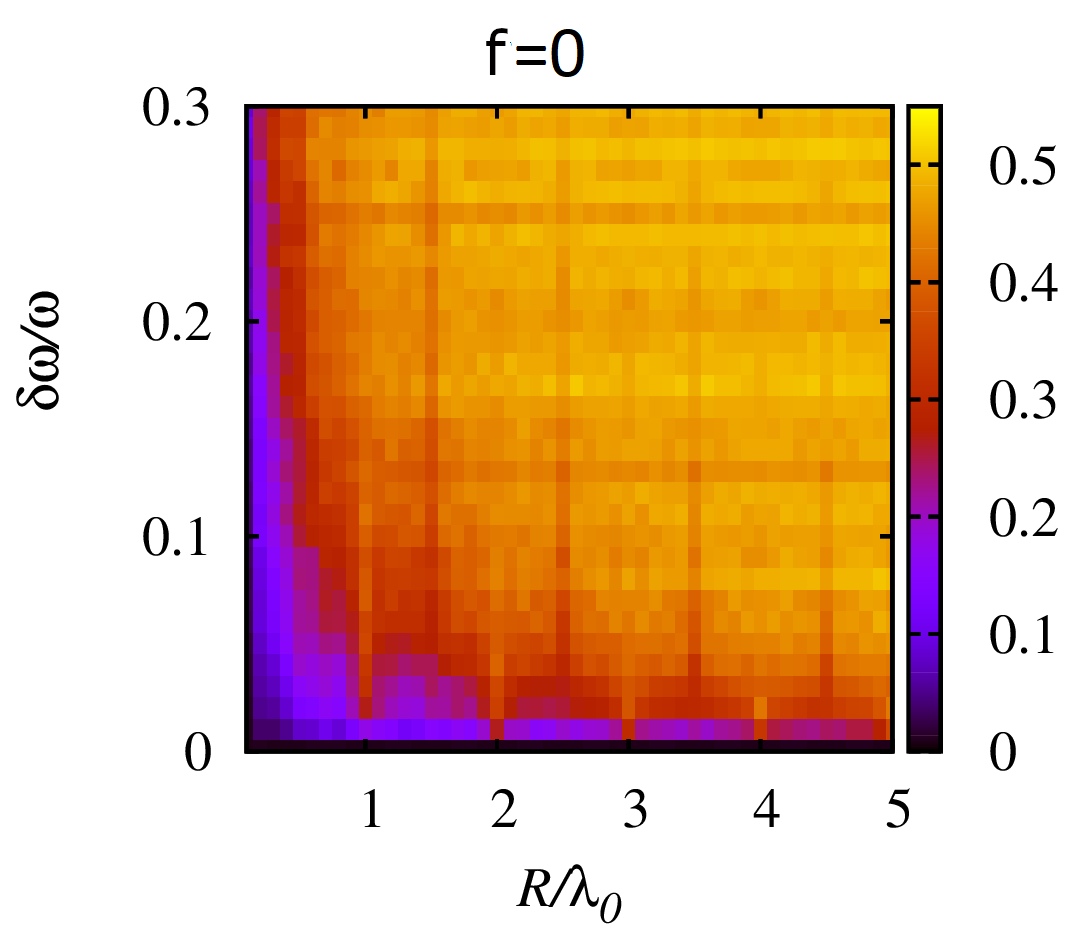}}
  \subfigure[]{%
  \includegraphics[width=0.45\textwidth]{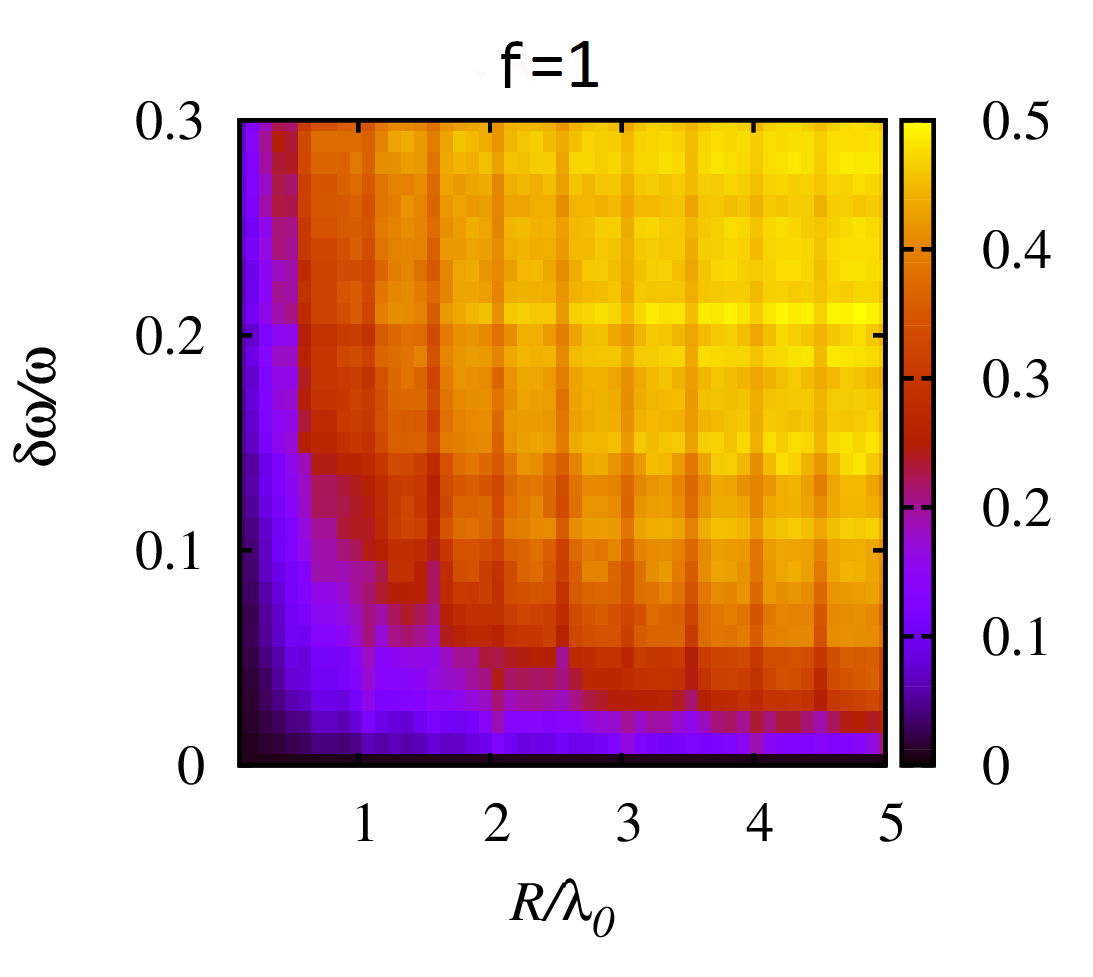}}
  \end{center}
      \caption{(a) The IPR averaged over all single excitation eigenstates as a function of $R/\lambda_0$ and disorder strength $\delta \omega/\omega$ for (a) $f=0$ and (b) $f=1$ respectively.  The system is described by Hamiltonian parameters: $\omega_0 = 0.2 \gamma$ and $N = 200$.}
      \label{Prob_WdisOmegaX}
  \end{figure*}

  \begin{figure*}
  \begin{center}
       \subfigure[]{%
  \includegraphics[width=0.75\textwidth]{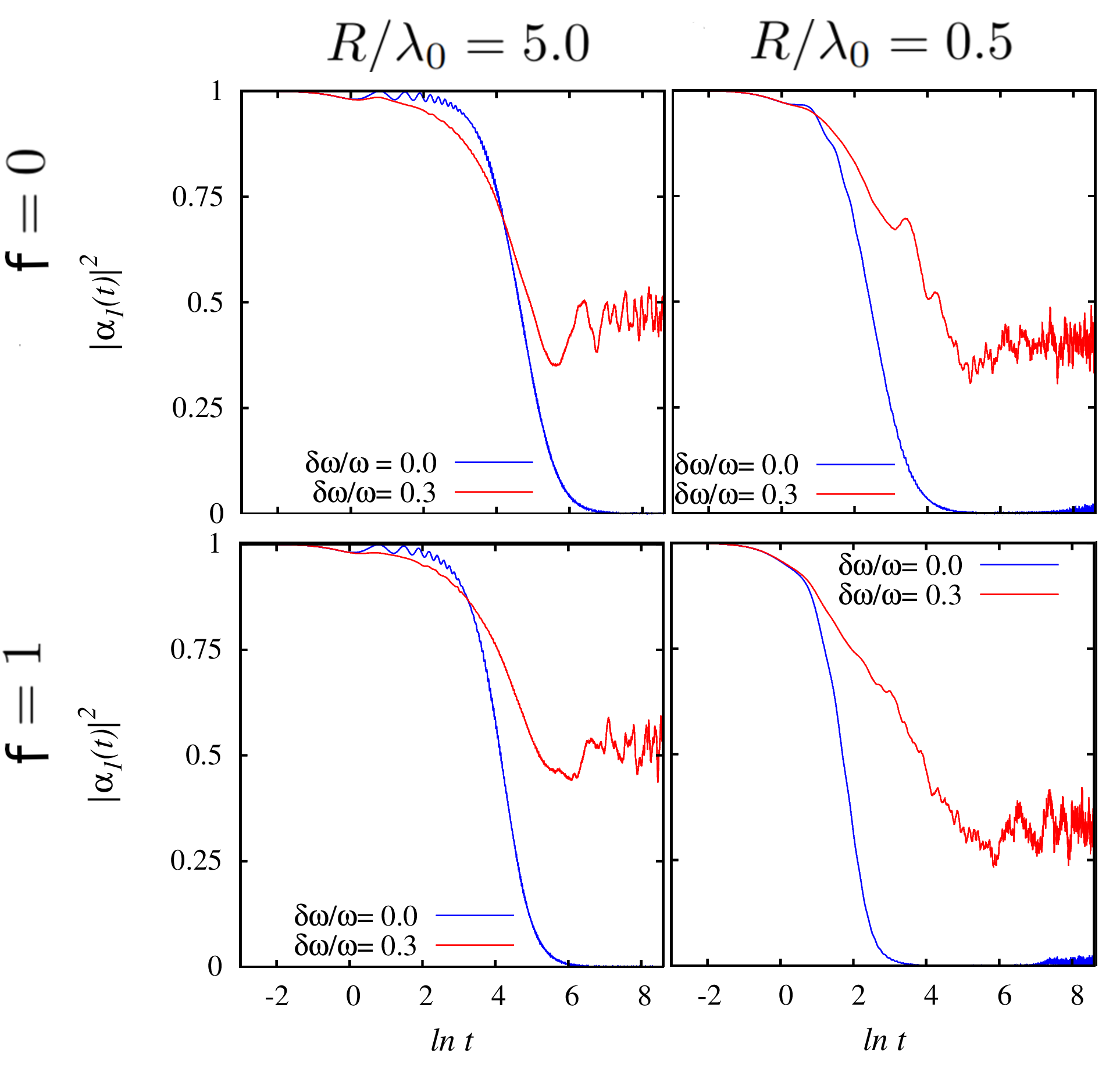}}
  \end{center}
      \caption{ The excitation probability of the initial site $|\alpha_1(t)|^2$ for $N=200$ as a function of  $\ln(t)$ (time in logarithmic scale) with and without disorder in $\omega$ in the presence ($f=1$) and absence ($f=0$) of dissipation. }
      \label{Prob_WdisOmega}
  \end{figure*}

\subsection{Entanglement entropy}
To quantify ergodicity and localization in the two-excitation sector, we use the bipartite entanglement entropy (EE) as a diagnostic. Given an eigenstate $|\psi_\alpha \rangle$ of the full Hamiltonian in the two-excitation sector, the reduced density matrix of a subsystem $A$ (taken as the sites $j = 1, . . . , N/2$) is defined as
\begin{eqnarray}
    \rho_A^{(\alpha)} = \text{Tr}_B |\psi_\alpha \rangle \langle \psi_\alpha|,
\end{eqnarray}
where $B$ denotes the complement of $A$. The corresponding EE reads
\begin{eqnarray}
    S_\alpha^{EE} (N) = -\text{Tr}_A \left(\rho^{(\alpha)}_A ln \rho^{(\alpha)}_A \right).
\end{eqnarray}
Since individual eigenstates may fluctuate strongly in their entanglement content, one typically considers the averaged EE, 
\begin{eqnarray}
    \langle S^{EE} (N) \rangle =\frac{1}{\mathcal{N}} \sum_{\alpha=1}^{\mathcal{N}} S^{EE}_{\alpha} (N),
\end{eqnarray}
where $\mathcal{N}$ is the total number of eigenstates in the two-excitation sector. This eigenstate-averaged EE captures the typical scaling
behavior: in the ergodic phase, $\langle S^{EE} (N) \rangle$ scales linearly with system size (volume law), while in the localized phase it saturates
to a constant (area law)~\cite{luitz2015many,luitz_prb_2016}.

We emphasize that this eigenstate-based definition differs from the EE obtained by propagating an arbitrary two-particle ini-
tial state $|\psi (0)\rangle$ under unitary dynamics, constructing $\rho_A (t) = \text{Tr}_B |\psi (t)\rangle \langle \psi (t)|$, and then computing $S^{EE} (t)$. The latter approach
provides time-dependent entanglement growth, while the former (eigenstate averaging) probes the statistical structure of eigen-
functions and their compliance with the ETH.

\section{Diagonal (energy) disorder}
\label{app:D}

We consider the disordered transition frequency strength around a mean transition frequency strength $\omega$. This is the energy disorder (also termed as \textit{diagonal disorder}). In a similar fashion as prevalent in position disorder, the diagonal disorder is introduced in the TLAs, resulting in the spread of the distribution of $\omega_0$ around the mean $\omega_m$ with a variance $\delta \omega_m$.

Fig.~\ref{Prob_WdisOmegaX} describes the variation of the IPR, averaged over all possible single-excitation eigenstates with $R/\lambda_0$ for different values of randomness. For both cases (a) and (b), the values of the IPR increase with increasing randomness, irrespective of the value of the interatomic distance. The contour demarcating the localized and the ergodic phase drops non-linearly and monotonically with increasing interatomic distance.

From Fig.~\ref{Prob_WdisOmega}, we plot the probability of the excitation at the first atom as a function of time in logarithmic scale for different interatomic distances, without and with dissipation. The plots connote that the probability decays to zero with time for a perfectly ordered chain, i.e., all the atoms with equal energies. On the other hand, in the presence of diagonal disorder, the probability saturates to a non-zero value, indicating localization in the system. For interatomic distances $R/\lambda_0$ equal to  $0.5$ and $5.0$, we observe that the asymptotic values of the probabilities saturate to 0.3 and 0.5, respectively, for the same randomness present in the system.

% \begin{figure}
%     \centering
%     \includegraphics[width=0.75\linewidth]{prob_density_pertb_1.pdf}
%     \caption{Caption}
%     \label{fig_pertb}
% \end{figure}

%\bibliography{lit-therm-coh}